\documentclass[letterpaper,twocolumn,10pt]{article}
\usepackage{usenix}

\usepackage{textcomp}
\usepackage{setspace}
\usepackage{latexsym,fancyhdr,url}
\usepackage{enumerate}
\usepackage[linesnumbered,ruled]{algorithm2e}
\usepackage{graphics}
\usepackage{xparse} %
\usepackage{xspace}
\usepackage{multirow}
\usepackage{csvsimple}
\usepackage{balance}
\usepackage{amsmath}
\usepackage{booktabs}
\usepackage{enumitem}
\usepackage{colortbl}
\usepackage{amsmath,amssymb,amsfonts}
\usepackage{fancyhdr}
\usepackage{makecell}
\usepackage{xcolor, eucal}
\usepackage{array}
\usepackage{subfigure}
\usepackage{graphicx}
\usepackage{caption}
\usepackage{tabularx}
\usepackage{dsfont}
\definecolor{gray0}{gray}{0.9}
\definecolor{mine}{RGB}{205, 232, 248} 
\usepackage{algpseudocode}

\usepackage{amsmath,amssymb,amsfonts}
\usepackage{graphicx}
\usepackage{textcomp}
\usepackage{booktabs}
\usepackage{hyperref}
\usepackage{multirow}
\usepackage{makecell}

\usepackage{amsmath,amsthm}
\usepackage{epsfig,endnotes}
\usepackage{xspace}
\usepackage{enumitem}
\usepackage{cleveref}
\usepackage{float}

\def \toolname{FETA-Pro}
\def \toolnamef{$\text{FETA\mbox{-}Pro}_{ft}$}
\def \toolnamem{$\text{FETA\mbox{-}Pro}_{mix}$}
\def \toolnameof{$\text{FETA\mbox{-}Pro}_{f}$}

\newcommand\rev[1]{{\color{black}  #1}}

\newtheorem{definition}{\bf Definition}[]
\newtheorem{theorem}{\bf Theorem}

\usepackage{filecontents}

\begin{document}

\date{}

\title{\Large \bf From Easy to Hard++: Promoting Differentially Private Image Synthesis Through Spatial-Frequency Curriculum}

\author{
{\rm Chen Gong\textsuperscript{$\dag$}}\\
University of Virginia
\and
{\rm Kecen Li\textsuperscript{$\dag$}}\\
University of Virginia
\and
{\rm Zinan Lin}\\
Microsoft Research
\and
{\rm Tianhao Wang}\\
University of Virginia
} %

\maketitle

\renewcommand{\thefootnote}{\fnsymbol{footnote}}
\footnotetext[2]{Equal contribution. Kecen works as a remote intern at UVA.}

\begin{abstract}
To improve the quality of Differentially private (DP) synthetic images, most studies have focused on improving the core optimization techniques (e.g., DP-SGD). 
Recently, we have witnessed a paradigm shift that takes these techniques off the shelf and studies how to use them together to achieve the best results. 
One notable work is DP-FETA, which proposes using `central images' for `warming up' the DP training and then using traditional DP-SGD. 

Inspired by DP-FETA, we are curious whether there are other such tools we can use together with DP-SGD. We first observe that using `central images' mainly works for datasets where there are many samples that look similar. To handle scenarios where images could vary significantly, we propose \toolname, which introduces {\it frequency features} as `training shortcuts.' The complexity of frequency features lies between that of spatial features (captured by `central images') and full images, allowing for a finer-grained curriculum for DP training. To incorporate these two types of shortcuts together, one challenge is to handle the training discrepancy between spatial and frequency features. To address it, we leverage the pipeline generation property of generative models (instead of having one model trained with multiple features/objectives, we can have multiple models working on different features, then feed the generated results from one model into another) and use a more flexible design.  Specifically, \toolname{} introduces an \textit{auxiliary generator} to produce images aligned with noisy frequency features. Then, another model is trained with these images, together with spatial features and DP-SGD. Evaluated across five sensitive image datasets, \toolname{} shows an average of 25.7\% higher fidelity and 4.1\% greater utility than the best-performing baseline, under a privacy budget $\epsilon = 1$. 
\end{abstract}

\section{Introduction}
\label{sec:intro}

Privacy-preserving image synthesis
aims to create artificial image data that retains the characteristics of real images, facilitating image data sharing within and between organizations while reducing privacy concerns~\cite{hu2023sok,zhang2021privsyn,sun2024netdpsyn}. Differentially private (DP) image synthesis~\cite{hu2023sok,gong2025dpimagebench,li2023PrivImage,dpsda,pe3,wang2025synthesize} provides a rigorous theoretical framework to quantify and limit privacy leakage from real image data using synthetic datasets. 

Most DP image synthesis methods rely on non-sensitive public resources released on open-source platforms like HuggingFace. 
They include pre-training synthesizers on public datasets~\cite{li2023PrivImage,dpldm,dplora} or leveraging pre-trained models~\cite{dpsda,pe3} to enhance synthetic performance. Tramèr et al.~\cite{position} note that public datasets may potentially compromise privacy. For example, the GPT-2 language model, pretrained on public web data, memorized the phone number of an individual named `Peter W.'~\cite{carlini2021extracting,wang2023decodingtrust}.
In addition, in reality, we may not find public datasets that are aligned well with sensitive datasets, and in such cases, these methods perform badly~\cite{gong2025dpimagebench}. Therefore, \emph{we aim at developing an effective DP
image synthesis method that does not rely on public resources.}

\noindent \textbf{Existing Methods.} 
Various works focus on developing DP Stochastic Gradient Descent (DP-SGD)~\cite{dpsgd} to improve DP image synthesis~\cite{dplora,dpldm}. Recently, we
witnessed a paradigm shift -- DP-FETA~\cite{dp-feta}  that uses `central images' to warm up synthesizers. The `central images' can be regarded as a `shortcut' to improve the training efficiency of DP-SGD~\cite{dp-feta}. Specifically, DP-FETA (1) extracts `central images' by averaging pixel values of aggregated images to warm up synthesizers, and (2) then fine-tunes synthesizers on original images using DP-SGD. DP-FETA achieves state-of-the-art performance in DP image synthesis without using public resources. However, 'central images' capture only simplistic characteristics, making them difficult to handle complex images and most effective for datasets where many samples are similar.

\noindent \textbf{Our proposal.} 
We are curious whether there are other such `training shortcuts' we can use together with DP-SGD, and refine them to enhance DP-SGD for DP image synthesis. To address the issue of the limited capacity of `central images,' we introduce frequency information as `training shortcuts.' This is inspired by previous works~\cite{crabbe2024time,li2025gaussmarker} that highlight the effectiveness of frequency features, and we use them as a complement to spatial features. As detailed in Section~\ref{subsec:moti}, frequency features provide richer information, like global structures and textures, about sensitive datasets. 
Overall, we propose \toolname{}, a spatial-frequency curriculum, which integrates frequency and spatial `training shortcut.' 

The next question is when to use the frequency training shortcut. Wang et al.~\cite{wang2021survey} suggest that the transition from easy to hard curriculum can benefit the completion of target tasks. As introduced in~\Cref{subsec:moti}, the complexity of frequency features falls between that of spatial features (captured by `central images,) and the full image, allowing for a more fine-grained curriculum for DP training. Thus, \toolname{} learns from the order of `spatial features $\to$ frequency features $\to$ images.' Referring to DP-FETA~\cite{dp-feta}, we use `central images' as spatial features, and extract frequency features using the Fourier Transform~\cite{dp-merf}. We list challenges and solutions when incorporating frequency features as a `training shortcut.'

\begin{itemize}[leftmargin=*]
\item \textit{Challenge I.} A key challenge arises in the training discrepancy
between the spatial and frequency features. To address this challenge, we propose learning spatial and frequency features on a unified representation. Specifically, \toolname{} first extracts frequency features and introduces noise to ensure DP. Subsequently, \toolname{} introduces an \textit{auxiliary generator} to learn to generate images aligned with the noisy frequency features. These synthetic images implicitly encode the frequency features, which are used to learn the underlying frequency features.

\item \textit{Challenge II.} For the design of auxiliary generators, diffusion models, which are used in \toolname{} as synthesizers, referring to previous works~\cite{dp-feta,li2023PrivImage}, are not suitable as auxiliary generators. 
Diffusion models learn via a forward diffusion process, incrementally introducing noise to a clean input~\cite{ddpm}. Determining the precise frequency features at each step of the diffusion process is challenging. Therefore, we use \textit{one-step} generators in the generative adversarial network (GAN)~\cite{gan}, which generates images only via one step, as the auxiliary generator. 
\end{itemize}

\begin{figure}[!t]
    \centering
    \setlength{\abovecaptionskip}{0pt}
    \includegraphics[width=1.0\linewidth]{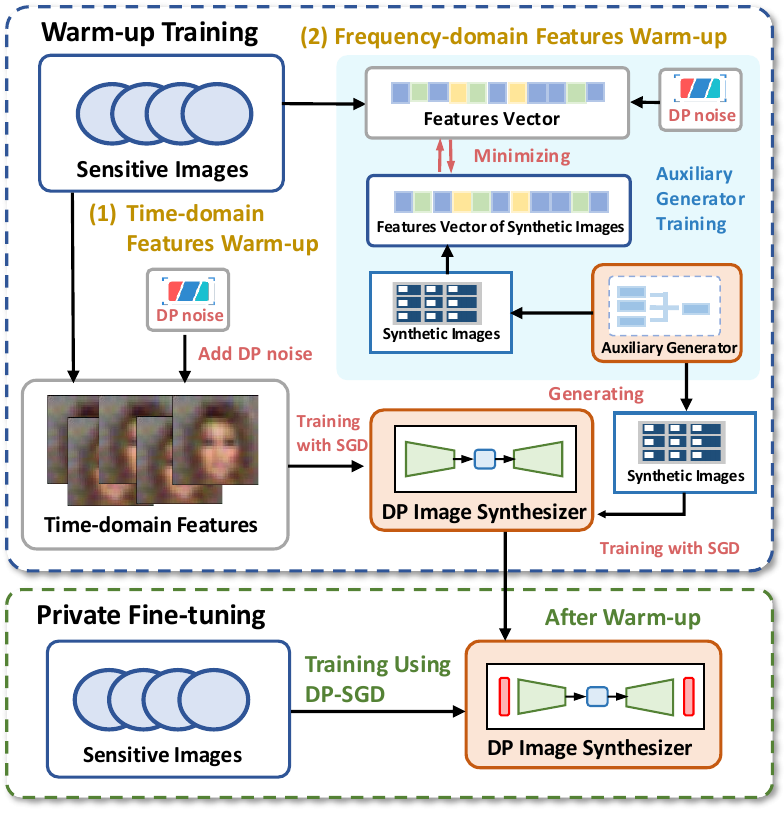}
    \caption{The framework of \toolname. During warm-up, \toolname{} extracts spatial features to train the synthesizer. Then, \toolname{} introduces auxiliary generators to generate images aligning the noisy frequency features, and training synthesizers on these synthetic images. Then, we train the warmed-up synthesizers on sensitive images using DP-SGD.}
    \label{fig:utility}
\end{figure}

Thus, instead of training one model on multiple features or objectives, \toolname{} leverages the pipeline generation property of generative models. This allows us to use a more flexible design by having multiple models work on different features. We discuss more about the challenges and solutions in~\Cref{subsubsec:chall}. Finally, the warm-up diffusion model is trained on the original sensitive images leveraging DP-SGD~\cite{dpsgd}. \Cref{fig:utility} presents the framework of \toolname.

\vspace{1mm}
\noindent \textbf{Evaluations.} 
We compare \toolname{} to seven baselines on five widely used sensitive datasets. Compared to the state-of-the-art method DP-FETA~\cite{dp-feta}, \toolname{} achieves 25.7\% higher fidelity and 4.1\% greater utility for synthetic images, under privacy budget $\epsilon = 1$. In particular, on the human face dataset {\tt CelebA}~\cite{celeba} and the medical dataset {\tt Camelyon}, \toolname{} enhances synthetic image quality by 20.2\% and 41.2\% in FID~\cite{fid}, and improves downstream classification task accuracy (Acc) by 7.7\% and 6.7\%, respectively, compared to DP-FETA~\cite{dp-feta}. Besides, the synthetic
performance of FETA-Pro converges more rapidly than that of baselines. These results show that \toolname{} incorporates frequency features as the shortcut, significantly improving the performance of DP image synthesis. Additionally, we study various privacy budget allocation strategies for spatial and frequency features in \toolname. We observe that allocation strategies highly affect synthetic image quality. Overall, we recommend allocating privacy budget cost ratios ordered as the complexity of features, like `spatial features < frequency features < images.'

\toolname{} also maintains good efficiency: Compared to DP-FETA~\cite{dp-feta}, 
for {\tt CIFAR-10}, \toolname{} incurs only a 0.3\% increase in training (approximately 0.06 hours).

\begin{table}[!t]
\footnotesize
    \centering
    \caption{The type of synthesizer used in existing methods. The first six methods listed above the horizontal line rely solely on sensitive data, and the other methods use public resources. } %
    \label{tab:SumOfMethods}
    \resizebox{0.49\textwidth}{!}{
    \begin{tabular}{l|cc}
    \toprule
    \textbf{Method} & \textbf{Training Methods} & \textbf{Type of Synthesizer}\\
    \hline
    DP-MERF~\cite{dp-merf} & Regular SGD & GAN \\
    DP-NTK~\cite{dp-ntk} & Regular SGD & GAN \\
    DP-Kernel~\cite{dp-kernel} & Regular SGD & GAN \\
    DP-GAN~\cite{dpgan} & DP-SGD & GAN \\
    DPDM~\cite{dpdm} & DP-SGD & Diffusion Model \\
    DP-FETA~\cite{dp-feta} & DP-SGD + Shortcut & Diffusion Model \\
    \rowcolor{gray0} \toolname{} (Ours) & DP-SGD + Shortcut & GAN \& Diffusion Model\\
    \hline
    PDP-Diffusion~\cite{dp-diffusion} & DP-SGD & Diffusion Model \\
    PE~\cite{dpsda} & - & APIs \\
    PrivImage~\cite{li2023PrivImage} & DP-SGD & Diffusion Model \\
    DP-LDM~\cite{dpldm} & DP-SGD & Diffusion Model \\
    DP-LoRA~\cite{dplora} & DP-SGD & Diffusion Model \\
    \bottomrule
\end{tabular}}
\end{table}

\vspace{1mm}
\noindent \textbf{Contributions.} We list our contributions as follows,
\begin{itemize}[leftmargin=*]
    \item This paper proposes \toolname, a spatial-frequency curriculum, that refines `training shortcuts'  by incorporating spatial and frequency features to warm up synthesizers.
    \item \Cref{tab:SumOfMethods} shows that current methods predominantly use a single type of model as the synthesizer. \toolname{} redefines DP image synthesis by integrating multiple synthesizers’ strengths (i.e., diffusion models and GANs), surpassing traditional single-synthesizer approaches to enhance performance and paving a path for further methods. %
    \item Comprehensive experiments present that \toolname{} accelerates DP diffusion model training while achieving state-of-the-art fidelity and utility across five image datasets, and only incurs minimal additional computer resources.
\end{itemize}

\section{Backgrounds}

\subsection{Differential Privacy}
\label{sub:dp}
Differential privacy (DP)~\cite{dp} quantifies the risk of individual privacy leakage within a dataset during data processing, serving as the gold standard for privacy preservation. We introduce the concept of DP as follows.

\begin{definition}[\textit{Differential Privacy}~\cite{dp}]
     A randomized mechanism $\mathcal{Q}$ satisfies $(\varepsilon, \delta)$-DP %
     if for any two neighboring datasets $D$ and $D'$, the following condition holds,
\begin{equation}\label{eq:dp}
    \Pr[\mathcal{Q}(D) \in \mathcal{O}] \leq e^\varepsilon \Pr[\mathcal{Q}(D') \in \mathcal{O}] + \delta.
\end{equation}
\end{definition}
\noindent Here, $\mathcal{O}$ represents a set of possible outputs of the algorithm $\mathcal{Q}$. $\mathcal{Q}$ is also referred to as the query function in the literature~\cite{zhang2021privsyn}. The parameters $(\varepsilon, \delta)$ quantify the privacy loss, with both being non-negative. A smaller $\varepsilon$ indicates stronger privacy protection, while $\delta$ represents the probability of failure, where a smaller $\delta$ reduces the likelihood that the privacy guarantees provided by $\varepsilon$ are violated~\cite{dp}. Datasets $D$ and $D'$ are considered neighboring if one can be obtained from the other by adding or removing a single image.

If an algorithm $\mathcal{Q}$ satisfies $(\epsilon,\delta)$-DP, any post-processing function $\mathcal{F}$, applied as $\mathcal{F} \circ \mathcal{Q}$ and $\mathcal{F}$ depending only on $\mathcal{Q}$'s output, incurs no additional privacy loss~\cite{dp}.

\vspace{1mm}
\noindent \textbf{DP-SGD.} In machine learning, a widely adopted method for achieving DP is DP-SGD~\cite{dpsgd}. This approach modifies standard SGD by computing gradients using Poisson-sampled mini-batches, clipping the gradients of the model's parameters, and adding Gaussian noise to the clipped gradients during training. Formally, we sample a sub-batch of images $D_s^{\text{sub}}=\{h_i\}_{i=1}^{B}$ from the sensitive dataset $D_s$ with sampling probability $q$.
Drawing on the implementation of DP-SGD~\cite{dpsgd}, we use Poisson sampling at each iteration, where the batch size ${B}$ is not fixed but follows an expected batch size ${B^*} = qN$. $N$ means the size of the sensitive dataset.
 The parameters $\theta$ of the synthesizer are updated via the following noisy gradient,
\begin{equation}
\label{eq:dpsgd}
    \eta \left( \frac{1}{{B^*}} \sum_{i = 1}^{{B}} \text{Clip}\left(\nabla {\mathcal{L}}(\theta, h_i), C\right) + \frac{C}{{B^*}} \mathcal{N}(0, \sigma^2 \mathbb{I}) \right),
\end{equation}
\noindent where $\mathcal{L}$ is the objective function of diffusion models, and $\eta$ is the learning rate and $\sigma^2$ is the variance of Gaussian noise. $\text{Clip}(\nabla {\mathcal{L}}, C) = \min\left\{1,\frac{C}{||\nabla {\mathcal{L}}||_2}\right\}\nabla \mathcal{L}$ clips the norm of gradient smaller than the hyper-parameter $C$.

This paper uses R\'{e}nyi DP (RDP)~\cite{sgm} to track the privacy loss, as detailed in~\Cref{app:supp_dp}.

\vspace{1mm}
\noindent \textbf{DP Image Synthesis.} %
The goal is to generate synthetic datasets that closely resemble real data while preserving the privacy of individual data points. This allows organizations to share synthetic images for various applications with reduced privacy concerns.

\Cref{tab:SumOfMethods} lists synthesizers used in existing methods. 
Prior work primarily focuses on using a single synthesizer type. Given the exceptional generative capabilities of diffusion models, recent methods increasingly emphasize their adoption~\cite{li2023PrivImage,dpldm,dplora}, while \toolname{} combines the strengths of multiple synthesizers to enhance synthetic performance.
\Cref{sec:related} presents more details of current methods.

\subsection{Diffusion Model}
\label{subsec:dm}

Diffusion models, as synthesizers, achieve the SOTA performance and are widely adopted in DP image synthesis~\cite{dp-diffusion,dpdm,dp-feta,li2023PrivImage,dpsda,pe3}. This work builds upon prior studies and uses diffusion models as synthesizers. Diffusion models~\cite{ddpm} operate through two key processes:
\begin{itemize}[leftmargin=*]
    \item The \emph{forward diffusion process}, which gradually adds noise to a clean image $h_0$, generating a sequence of increasingly noisy images $\{h_t\}_{t=1}^T$ until it approximates pure random noise, where $T$ denotes the total number of noising steps.
    \item The \emph{reverse diffusion process}, which progressively denoises random noise to reconstruct a clean image.
\end{itemize}
In the forward diffusion process, the transition between consecutive noisy images, denoted $p(h_t | h_{t-1})$, follows a multidimensional Gaussian distribution, $
p(h_t | h_{t-1}) = \mathcal{N}\left( h_t; \sqrt{1 - \beta_t} h_{t-1}, \beta_t \mathbb{I} \right)$,
where $\beta_t$ is a hyper-parameter controlling the noise variance at each step. We define $\bar{\alpha}_t = \prod_{s=1}^t (1 - \beta_s)$. Consequently, the likelihood of a noisy image $x_t$ given the clean image $h_0$ is,
$p(h_t | h_0) = \mathcal{N}\left( h_t; \sqrt{\bar{\alpha}_t} h_0, (1 - \bar{\alpha}_t) \mathbb{I} \right)$. This enables direct sampling of $h_t$ from $h_0$ in closed form, ${h_t} = \sqrt {{{\bar \alpha }_t}} {h_0} + e_t\sqrt {1 - {{\bar \alpha }_t}} ,e_t\sim \mathcal{N}\left( {0,\mathbb{I}} \right).$
The objective of diffusion models is to train a neural network to predict the noise added at each step~\cite{ddpm}:
\begin{equation}
\label{eq:L_DM}
\mathcal{L} = \mathbb{E}_{h_0 \sim D, t \sim U\{1, T\}, e_t \sim \mathcal{N}(0, \mathbb{I})} \left\| e_t - e_\theta(h_t, t) \right\|_2^2
\end{equation}
where $D$ is the dataset of clean images, and $e_\theta(h_t, t)$ is a denoising network parameterized by $\theta$. $U\{1,T\}$ denotes a discrete uniform distribution over the integers from 1 to $T$. This network learns to predict the noise $\epsilon_t$ in a noisy image $h_t$ at step $t$. Once trained, $\epsilon_\theta$ enables the generation of clean images by denoising random Gaussian noise.

\begin{figure}[!t]
    \centering
    \setlength{\abovecaptionskip}{0pt}
    \includegraphics[width=0.98\linewidth]{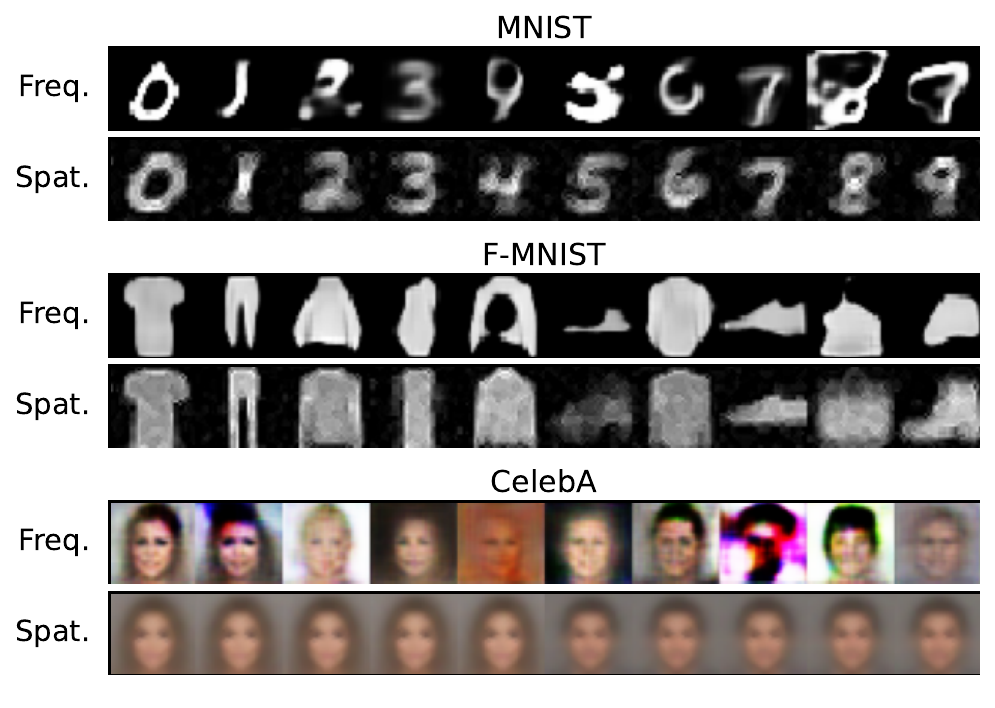}
    \vspace{1.5mm}
    \caption{Sampled images generated leveraging frequency (`Freq.') and spatial (`Spat.') domain features under $\epsilon=1$.}
    \label{fig:time_freq}
\end{figure}

\subsection{Frequency Domain Features of Images}
\label{subsec:theo}

Fourier Transform (FT) is a tool for analyzing images in the frequency domain. For a continuous image denoted as a function $I(x, y)$ over a spatial domain $[0, l] \times [0, w]$, where $l$ and $w$ represent height and width, the FT transforms the image into frequency components $F(u, v)$, 
\begin{equation}
    F(u, v) = \int_{0}^{g} \int_{0}^{w} I(x, y) e^{-j 2\pi (ux + vy)} \, dx \, dy,
\end{equation}
where $u, v$ are continuous frequency variables representing spatial frequencies in horizontal and vertical directions~\cite{jahne2005digital}.

\subsection{Training Shortcuts}
\label{subsec:training_shortcuts}
DP-FETA introduces using `central image' to augment DP-SGD, which we view as a shortcut for training. In this paper, we provide a formal description of this concept.

`Training shortcuts' are simpler representations of a sensitive dataset, $D_s$, that (1) are beneficial for the synthesizer to learn complex images and (2) only need minimal privacy loss to obtain, than the original, complex images. This knowledge is formally expressed as $F(D_s)$, where $F(\cdot)$ denotes the knowledge extraction method. Then, the synthesizer, $e$, is trained with DP to learn the mapping from simpler knowledge to the original data, following the order of `$e \to F(D_s) \to D_s$.'

\section{Methodology}
\label{sec:methodology}
This section details the spatial-frequency curriculum, \toolname, including the motivation and technical details.

\begin{table}[!t]
\footnotesize
    \centering
    \caption{The entropy and texture complexity of images generated leveraging frequency and spatial features. Higher values indicate greater image complexity. `Text. Comp.' is the abbreviation for texture complexity.}
    \label{tab:metric_s_f_o}
    \resizebox{0.48\textwidth}{!}{
    \begin{tabular}{l|cc|cc}
    \toprule
    \multirow{2}{*}{\textbf{Feature}} & \multicolumn{2}{c|}{{\tt CIFAR-10}} & \multicolumn{2}{c}{{\tt CelebA}} \\
    \cline{2-5}
     & Entropy & Text. Comp. & Entropy & Text. Comp. \\
    \hline
    Spatial & 0.522 & 1.228 & 0.857 & 1.857\\
    Frequency & 1.064 & 1.560 & 1.192 & 1.966\\
    Original Images & 1.243 & 2.169 & 1.336 & 2.363\\
    \bottomrule
\end{tabular}
}
\end{table}

\subsection{Motivation}
\label{subsec:moti}
\noindent DP-FETA uses `central images' (serve as basic spatial features) extracted from sensitive data as a training shortcut. However, these features capture only coarse and simplistic characteristics, making it difficult to handle complex datasets. For example, as shown in Table~\ref{tab:rq1}, the FID on the {\tt Camelyon} dataset is only 52.8 under privacy budget $\epsilon=1$.

Previous works~\cite{crabbe2024time,li2025gaussmarker} have shown that frequency features, which capture global structures and textures, are effective for various tasks. This inspires us to integrate frequency features as a training shortcut from the two perspectives.
\begin{itemize}[leftmargin=*]
\item \textit{Complementing spatial features.} \Cref{fig:time_freq} displays synthetic images generated by synthesizers that respectively use frequency and spatial features under $ \epsilon = 1$. This figure shows that spatial features capture the shape and color details of images, while frequency features emphasize texture characteristics and more clearly capture edge variations.
\item \textit{Fine-grained curriculum.} Wang et al.~\cite{wang2021survey} advocate for fine-grained curricula to enhance task effectiveness. As presented in~\Cref{tab:metric_s_f_o}, we observe that frequency features possess a complexity between spatial features and full images. We provide more details of `entropy' and `complexity' in~\Cref{apsubsec:MetricDetails}. This observation forms the basis for our curriculum design in DP image synthesis. \toolname{} thus orders features by complexity, progressing from `spatial features' to `frequency features' and finally to `images.'
\end{itemize}

\Cref{sec:rq2} presents that integrating these two types of features achieves better synthetic performance than using either one in isolation. Besides,~\Cref{sec:rq2} validates that \toolname{} achieves the best synthetic performance compared to learn spatial and frequency features concurrently and learn through prioritizing frequency features over spatial features.

\vspace{1mm}

\subsection{Feature Extraction}
\label{subsec:frequent_feature}

This section explains how to extract frequency and spatial features from sensitive datasets under DP as training shortcuts.

\begin{algorithm}[!t]
       \caption{Feature Extractions in \toolname.}
      \label{alg:feta-pro-feature}
       \SetKwInOut{Input}{Input}
      \SetKwInOut{Output}{Output}
      \SetKwFunction{SpaticalFeatureQuery}{SpaticalFeatureQuery}
      \SetKwProg{Fn}{Function}{:}{}
    \Input{Sensitive dataset $D_s$ with estimated size $N^\ast$; number of central images $N_t$, batch size $B_t$; pre-defined vector length $K$; the height and width of images $l$ and $w$.}
    \tcp{Spatial feature extraction}
    Init noisy central image dataset $D_s^t = \varnothing$; \\
    \While{$\text{len}\left(D_s^t\right) < N_t$}{
        Sample subset $D_{s}^{\text{sub}}=\{h_i\}_{i=1}^{B_t}$ from $D_s$;\\
        Calculate noisy central image $\Tilde{h}$ using~\Cref{eq:meanImage};\\
        $D_s^t = D_s^t \cup \{\Tilde{h}\}$;\\
    }
    \tcp{Frequency feature extraction}
    \For{$h_i \in D_s$ \label{line:fe_begin}}{
    Vectorize $h_i \in \mathbb{R}^{l\times w}$ to $h_i \in \mathbb{R}^{d}$, where $d = l\times w$; \\
        \For{$j = 1,2,\cdots,K/2$}{
            $\omega_j \in \mathbb{R}^d \sim \mathcal{N}(0,\mathbb{I})$; \\
            $\phi_j(h_i)=\sqrt{2/K} \cos{(\omega_j^{\top}h_i)}$; \\
            $\phi_{j+K/2} (h_i)=\sqrt{2/K} \sin{(\omega_j^{\top}h_i)}$; \\
        }
        $\phi(h_i) = {\left(\phi_1(h_i), \ldots, \phi_K(h_i)\right)}^{\top}$,
    }
    Calculate mean frequency features: $\mu = \frac{1}{N^*} \sum_{i=1}^N \phi(h_i), \mu \in \mathbb{R}^K$; \label{line:fe_end} \\ %
    Add noise to $\mu$ for DP using~\Cref{eq:freq_noise} and obtain $\Tilde{\mu}$; \\
    \Output{Spatial and frequency feature, $D_s^t$ and $\Tilde{\mu}$.}
\end{algorithm}

\noindent \textbf{Frequency Features Shortcuts.} Random Fourier Features~\cite{rahimi2007random,Zhang_2025_CVPR,dp-merf} are preferred over the traditional Fourier Transform for image feature extraction in image processing due to their computational efficiency and ability to represent arbitrary dimensions of images as fixed-length vectors~\cite{rahimi2007random}. We refer to implementations in previous works~\cite{dp-merf}.
Given an image $h$, its random Fourier features are given by, $\phi(h) = {\left(\phi_1(h), \ldots, \phi_K(h)\right)}^{\top},$ where $K$ is the pre-defined feature dimension, and each coordinate is calculated by,
\begin{equation}
\left\{
    \begin{split}
    & \phi_j(h)=\sqrt{2/K} \cos{(\omega_j^{\top}h)}\\
    & \phi_{j+K/2}(h)=\sqrt{2/K} \sin{(\omega_j^{\top}h)}&,
    \end{split}
\right.
\label{eq:rff}
\end{equation}
\noindent where $j=\{1,\ldots, K/2 \}$, and $\omega_j \in \mathbb{R}^d$ is a random frequency vector, with each component independently drawn from a standard normal distribution. The $d = l\times w$ is the dimension of the vectorized image $h$. 
$l$ and $w$ represent the height and width of images. The term $\omega_j^\top h$ is the inner product $\langle \omega_j, h \rangle$, projecting the image onto the random frequency $\omega_j$, enabling the construction of cosine and sine features to capture frequency features~\cite{rahimi2007random,jahne2005digital}. 

For a sensitive image dataset $D_s = \{ h_i \}_{i=1}^N$ with $N$ images, the mean frequency features are, $\mu = \frac{1}{N^\ast} \sum_{i=1}^N \phi(h_i)$, where $N^*$ approximates dataset size $N$ as introduced in Appendix~\ref{apsec:proof}. 
\begin{equation}
\label{eq:freq_noise}
    \tilde{\mu}=\mu + \mathcal{N}(0, \sigma_f^2\Delta_f^2\mathbb{I}),
\end{equation}
\noindent where $\sigma_f^2$ is a hyper-parameter describing the scale of Gaussian noise, and the global sensitivity $\Delta_{f}=1/N^\ast$. \Cref{the:frequency} proves that this process ensures DP.

\begin{theorem}
\label{the:frequency}
    The query of frequency features of $D_s$ has global sensitivity $\Delta_{f}=1/N^\ast$. For any $\alpha>1$, incorporating noise $\mathcal{N}\left(0,{\sigma_f^2 \Delta_f^2} \mathbb{I} \right)$ into the mean random Fourier feature $\mu$ makes the query results satisfy $\left(\alpha, \gamma \right)$-RDP for some $\gamma$. 
\end{theorem}

We detail $\left(\alpha, \gamma \right)$-RDP in~\Cref{app:supp_dp}. Besides, we provide proof of~\Cref{the:frequency} in Appendix~\ref{apsec:proof}.

\vspace{1.0mm}
\noindent \textbf{Spatial Features Shortcuts.}  We borrow the approach of extracting spatial features from DP-FETA~\cite{dp-feta}, which uses central images as spatial shortcuts.
We first sample a subset of $B_t$ sensitive images, $D_s^{\text{sub}} = \{ h_i \}_{i=1}^{B_t}$, from the sensitive dataset $D_s=\{h_i\}_{i=1}^N$, consisting of $N$ images, using Poisson subsampling with probability $ q_t $.  %
The batch size $B_t$ varies and is not fixed in each step. We only have access to the expected batch size,  $B_t^* = q_t N^*$, where $N^*$ is the estimated dataset size. Each sensitive image is then clipped to ensure a bounded $L_2$ norm, defined as $h_i^c = \min\left\{1, \frac{C_t}{\|h_i\|_2} \right\} \cdot h_i$, where $C_t$ is a hyper-parameter, guaranteeing $\|h_i^c\|_2 \leq C_t $. The central image is computed as $ h^\text{spat}=\frac{1}{B_t} \sum_{i=1}^{B^*_t} h_i^c $. Then, Gaussian noise is added to the central images to obtain final spatial features $\Tilde{h}$,
\begin{equation}
    \label{eq:meanImage}
    \frac{1}{{B_t^\ast}} \sum_{i=1}^{B_t} {\min\left\{1,\frac{C_t}{||h_i||_2}\right\}\cdot h_i} + \mathcal{N}\left(0,{\sigma_t^2 \Delta_\text{spat}^2} \mathbb{I} \right),
\end{equation}
\noindent where $\sigma_t^2$ is a hyper-parameter that controls the scale of the noise, and $\Delta_\text{spat}=C_t/B_t^\ast$. We can prove that this ensures DP.

\begin{algorithm}[!t]
       \caption{The workflow of \toolname.}
      \label{alg:feta-pro}
       \SetKwInOut{Input}{Input}
      \SetKwInOut{Output}{Output}
      \SetKwProg{Fn}{Function}{:}{}
    \Input{Diffusion model $e_\theta$ parameterized with $\theta$; sensitive dataset $D_s$; spatial and frequency feature, $D_s^t$ and $\Tilde{\mu}$;  auxiliary generator $G_{\theta'}$ parameterized with $\theta'$; training epoch of auxiliary generator $M$; training batch size of auxiliary generator $B_f$; the number of images sampled from the auxiliary generator $N_f$.}
    \tcp{Spatial domain warm-up}
    Train the synthesizer $e_\theta$ on $D^t_s$ using~\Cref{eq:L_DM}; \\
    \tcp{Frequency domain warm-up}
    Init $m=0$; \\
    \While{$m \leq M$}{
        Leverage $G_{\theta'}$ to generate images $D^{\text{tmp}} = \{h_i\}^{B_f}_{i=1}$; \\
        Extracting frequency features $\mu^s(G_{\theta'})$ from $D^{\text{tmp}}$ using~\Cref{line:fe_begin}-\Cref{line:fe_end} in~\Cref{alg:feta-pro-feature}; \\
        Update $\theta'$ by minimizing: $\mathcal{L}(\theta') = \|\tilde{\mu} - \mu^s(G_{\theta'})\|_2$; \\
        $m = m + 1$; \\
    }
    Leverage trained $G_{\theta'}$ to generate images $D^f_s = \{h_i\}^{N_f}_{i=1}$;\\
    Train the synthesizer $e_\theta$ on $D^f_s$using~\Cref{eq:L_DM}; \\
    \tcp{Private Fine-tuning}
    Fine-tuning the synthesizer $e_\theta$ on $D_s$ using DP-SGD; \\
    \Output{Well-trained DP image synthesizer $e_\theta$.}
\end{algorithm}

\begin{theorem}
\label{the:meanImage}
    The query of spatial features $h^\text{spat}$ has global sensitivity $\Delta_\text{spat}=C_t/B_t^\ast$. For any $\alpha>1$, incorporating noise $\mathcal{N}\left(0,{\sigma_t^2 \Delta_\text{spat}^2} \mathbb{I} \right)$ into the spatial features $h^\text{spat}$ makes the query results satisfy $\left(\alpha, \gamma \right)$-RDP for some $\gamma$. 
\end{theorem}

\noindent We provide the proof of~\Cref{the:meanImage} in Appendix~\ref{apsec:proof}. %
Repeating the above process $N_t$ times, we can obtain the noisy central image dataset $D_s^t = \{\Tilde{h}_i\}_{i=1}^{N_t}$ as the spatial features. \toolname{} uses RDP composition to analyze the privacy cost of extracted spatial and frequency features. We elaborate on the privacy analysis in~\Cref{subsec:privacy_analy}.

\Cref{alg:feta-pro-feature} presents the workflow of extracting spatial and frequency features from sensitive datasets under DP.

\vspace{1mm}
\noindent \textbf{Features Extraction for Labeled Data.} When data labels are available--a common scenario in image generation tasks--we can partition the sensitive dataset $D_s$ into multiple disjoint subsets based on labels. Specifically, we categorize $D_s$ into $C$ disjoint subsets, $D_s = [D_s^1, D_s^2, \cdots, D_s^C]$, based on their labels, where $C$ is the number of categories. Then, we extract spatial and frequency features for each subset. By the parallel composition property~\cite{dp}, processing disjoint subsets incurs no additional privacy cost, yielding the same privacy guarantees as if the entire dataset were processed without partitioning. 

For frequency features extraction, we first calculate the random Fourier features of each image and aggregate them in each subset $[D_s^1, D_s^2, \cdots, D_s^C]$. Then, we obtain the mean frequency features, $\mu = [\mu_1,\mu_2,\cdots,\mu_C]$, where $\mu_i = \frac{1}{m} \sum_{h\in D_s^i} \phi(h)$ and $\mu \in \mathbb{R}^{K\cdot C}$. $K$ is the pre-defined feature dimension. Due to the disjoint nature of the subsets, the privacy guarantee remains the same as that in~\Cref{the:frequency}.

\begin{theorem}
\label{the:frequency-disjoint}
    Consider partitioning the dataset as $D_s = [D_s^1, D_s^2, \cdots, D_s^C]$, where the subsets are disjoint. By applying~\Cref{eq:freq_noise} to each subset, we generate noisy frequency features. The combined features have a dimension of $K\cdot C$, where $K$ is the predefined feature dimension and $C$ is the number of subsets. Because the subsets are disjoint, this mechanism has the same $\left(\alpha, \gamma \right)$-RDP guarantee as that in~\Cref{the:frequency}.
\end{theorem}

The proof of~\Cref{the:frequency-disjoint} is provided in~\Cref{apsec:proof}. We then obtain the target frequency features using~\Cref{eq:freq_noise}. For spatial features, when sampling a batch of images to query a central image, we ensure that all sampled images are from a single subset. Similarly, the query of spatial features keeps consistent sensitivity as presented in~\Cref{the:meanImage}.

\subsection{Private Training}
\label{subsec:private_training}

This section first introduces challenges in private training and then elaborates on technical details.

\subsubsection{Challenges and Solutions}
\label{subsubsec:chall}
We present the challenges of incorporating frequency features as training shortcuts and solutions in \toolname{} as follows.

\begin{itemize}[leftmargin=*]
    \item \textbf{The discrepancy between spatial and frequency features poses a challenge for their learning together. } Spatial and frequency features exist in different domains; spatial features capture localized patterns in the data’s original representation, while frequency features, such as those from random Fourier transforms, represent global oscillatory characteristics in a transformed space.~\cite{sonka2013image}. %
    It is unclear how to train synthesizers on these two distinct domains together.

    To solve this challenge, we hope that synthesizers learn these features on a unified representation. Since diffusion models primarily learn in the spatial domain~\cite{ddpm,he2023diffusion}, we convert frequency feature learning to this domain.  Specifically, \toolname{} first extracts frequency features from sensitive images under DP. An \textit{auxiliary generator} is then trained to produce images that align with these noisy frequency features. We consider these synthetic images involving frequency features of sensitive images. 
    \item \textbf{Using diffusion models (the model of synthesizers) as the auxiliary generator is not suitable.} Diffusion models learn via a forward diffusion process, incrementally introducing noise to clean inputs~\cite{ddpm}. The specific frequency features of sensitive images at each step of the diffusion process remain unknown. We only have access to the frequency features of the clean sensitive images. Thus, multiple-step generation synthesizers (i.e., diffusion models) are suboptimal.   

    We adopt the \textit{one-step} generator~\cite{gan}, which generates images in a single step, as the auxiliary generator. The auxiliary generator learns to create images that approximate the noisy frequency features. These images, which incorporate frequency features, are then used by the main generator to warm up the diffusion model.
\end{itemize}

This design combines the pipeline generation property of different generative models, enabling different models to handle specific features and resulting in a more flexible framework for DP image synthesis. \Cref{tab:diffusion_frequnecy_ab} of~\Cref{sec:rq1} presents that learning frequency features directly using synthesizers without auxiliary generators and using diffusion models as auxiliary generators achieves inferior performance compared to our proposed \toolname{}.

\subsubsection{Technical Details}
\label{subsec:tech_details}

\vspace{1mm}
\noindent \textbf{Warm Up.} The spatial features consist of a noisy central image dataset $D_s^t$. We can first directly train the diffusion model on this dataset for warm-up. 

As we discussed in~\Cref{subsubsec:chall}, we introduce an auxiliary generator to learn to generate images aligning with the noisy frequency features $\Tilde{\mu}$ from the sensitive images $D_s$. Then, synthetic images from the auxiliary generator, which involve the frequency features of sensitive datasets, are used to further warm up the diffusion models. Specifically, we employ the generator from a GAN, which produces images from a random Gaussian vector $z$ in a single step, as the auxiliary generator. In each train epoch, we use a vector set $\{z_i\}_{i=1}^{B_f}$ to generate a synthetic image dataset $D^{\text{tmp}} = \{h_i\}_{i=1}^{B_f} = \{G_{\theta'}(z_i)\}_{i=1}^{B_f}$, where $B_f$ refers to the size of synthetic images and $\theta'$ is the network parameter. Referring to~\Cref{eq:rff}, we can obtain the frequency features of $D^{\text{tmp}}$, $\mu^s = \frac{1}{{B_f}}\sum_{i=1}^{B_f} \phi(h_i), h_i \in D^{\text{tmp}}$. We hope that the $\mu^s$ are close to the noisy features of sensitive datasets $D_s$, so the objective of the auxiliary generator is,
\begin{equation}
\label{eq:mmd}
    \mathcal{L}(\theta') = \|\tilde{\mu} - \mu^s(G_{\theta'})\|_2
\end{equation}
Note that we only borrow the generator in GAN. %
Unlike traditional GAN training, which relies on a critic to provide an adversarial loss to indirectly guide the generator~\cite{wgan,gan}, 
our critic (\Cref{eq:mmd}) is explicitly defined. This makes the training
more efficient and avoids the training instability issue in traditional GANs~\cite{wgan,dp-merf,lin2021spectral}.

\vspace{1mm}
\noindent \textbf{DP-SGD Fine-tuning.} Then, we fine-tune the warmed-up diffusion model on the original sensitive images to learn more complex features of the images. To achieve DP, we add Gaussian noise to the clipped training gradients and use the noisy gradient to update the model parameters following standard DP-SGD~\cite{dpsgd}, as introduced in~\Cref{sub:dp}. \Cref{alg:feta-pro} summarizes the workflow of \toolname{}.

\subsection{Privacy Analysis}
\label{subsec:privacy_analy}

In \toolname, three processes consume the privacy budget: (1) querying the frequency features, (2) querying the central images for spatial features, and (3) fine-tuning the warm-up diffusion model on the sensitive dataset using DP-SGD. According to Renyi DP (RDP)~\cite{rdp}, these three processes satisfy $(\alpha,\gamma_t)$-RDP, $(\alpha,\gamma_f)$-RDP and $(\alpha,\gamma_d)$-RDP, respectively. Specifically, $(\alpha,\gamma_t)$ is determined by the noise scale $\sigma_t$, and we use the whole sensitive dataset. $(\alpha,\gamma_f)$ is determined by the number of central images $N_t$, sample ratio $q_t$ and noise scale $\sigma_t$. $(\alpha,\gamma_d)$ is determined by the fine-tuning iteration $t_d$, sample ratio $q_d$ and noise scale $\sigma_d$~\cite{dpsgd}. Each process of \toolname{} can be viewed as a Sub-sampled Gaussian Mechanism (SGM)~\cite{sgm}. According to the RDP composition theorem~\cite{rdp}, \toolname{} satisfies $(\alpha,\gamma_t+\gamma_f+\gamma_d)$-RDP. We introduce the concept and property of RDP in~\Cref{app:supp_dp}.

To make \toolname{} satisfy a given $(\varepsilon,\delta)$-DP, we determine privacy parameters following three steps: (1) We set the number of central images $N_t$, sample ratio $q_t$ and noise scale $\sigma_t, \sigma_f$ to obtain the RDP costs $(\alpha, \gamma_t)$ and $(\alpha, \gamma_f)$. (2) We fix the fine-tuning iterations $t_d$ and sample ratio $q_d$, and then the RDP cost of DP-SGD is a function of noise scale $\sigma_d$ as $\left(\alpha, \gamma_d(\sigma_d)\right)$. (3) We try different $\sigma_d$ to obtain the corresponding $(\varepsilon,\delta)$-DP~\cite{rdp}, until meeting the given privacy budget. \Cref{apsubsec:privacy_allocation} discusses the privacy calculation in \toolname. \Cref{apsubsec:PRV} presents results under another privacy accounting method, Privacy Random Variable (PRV)~\cite{PRV}.

\begin{table}[!t]
\footnotesize
    \centering
    \caption{The data split and number of categories of sensitive image datasets used in our experiments.}
    \label{tab:datainfo}
    \resizebox{0.48\textwidth}{!}{
    \begin{tabular}{l|ccccc}
    \toprule
    \textbf{Sensitive Datasets} & \textbf{Training} & \textbf{Validation} & \textbf{Test} & \textbf{Category} \\
    \hline
    {\tt MNIST} & 55,000& 5,000 & 10,000 & 10\\
    {\tt F-MNIST} & 55,000& 5,000 & 10,000 & 10\\
    {\tt CIFAR-10} & 45,000& 5,000 & 10,000 & 10\\
    {\tt CelebA} & 162,770 & 19,867 & 19,962 & 2\\
    {\tt Camelyon} & 302,436 & 34,904  & 85,054 & 2\\
    \bottomrule
\end{tabular}
}
\end{table}

\section{Experimental Setup}
\label{sec:setup}

\noindent \textbf{Baselines.} \toolname{} achieves DP image synthesis without relying on public resources, such as public datasets or pre-trained models. Therefore, we select baselines under the same constraint, including DP-MERF~\cite{dp-merf}, DP-NTK~\cite{dp-ntk}, DP-Kernel~\cite{dp-kernel}, GS-WGAN~\cite{g-pate}, DP-GAN~\cite{dpgan}, DPDM~\cite{dpdm}, and DP-FETA~\cite{dp-feta}. The implementations use the open-source DP image synthesis benchmark, DPImageBench \cite{gong2025dpimagebench}.\footnote{\url{https://github.com/2019ChenGong/DPImageBench}} We provide more baselines details in the~\Cref{apsubsec:baselines}.

\begin{table*}[!t]
\renewcommand{\arraystretch}{1.1}
\setlength{\tabcolsep}{5.5pt}
\small
    \centering
    \caption{FID and Acc (\%) of \toolname{} and seven baselines on {\tt MNIST}, {\tt F-MNIST}, {\tt CIFAR-10}, {\tt CelebA} and {\tt Camelyon} with $\varepsilon=\{1,10\}$. The best and second-best values are highlighted in bold and underlined in each column.}
    \label{tab:rq1}
    \resizebox{1.0\textwidth}{!}{
    \begin{tabular}{l|rc|rc|rc|rc|rc|rc|rc|rc|rc|rc}
    \toprule
    \multirow{3}{*}{\textbf{Method}} & \multicolumn{10}{c|}{$\varepsilon=1$} & \multicolumn{10}{c}{$\varepsilon=10$}\\
    \Xcline{2-21}{0.5pt}
    & \multicolumn{2}{c|}{{\tt MNIST}} & \multicolumn{2}{c|}{{\tt F-MNIST}} & \multicolumn{2}{c|}{{\tt CIFAR-10}} &  \multicolumn{2}{c|}{{\tt CelebA}} & \multicolumn{2}{c|}{{\tt Camelyon}} & \multicolumn{2}{c|}{{\tt MNIST}} & \multicolumn{2}{c|}{{\tt F-MNIST}} & \multicolumn{2}{c|}{{\tt CIFAR-10}} & \multicolumn{2}{c|}{{\tt CelebA}} & \multicolumn{2}{c}{{\tt Camelyon}} \\
    \Xcline{2-21}{0.5pt}
     & \centering FID & Acc & FID & Acc & FID & Acc & FID & Acc & FID & Acc & FID & Acc & FID & Acc & FID & Acc & FID & Acc & FID & Acc\\
    \hline
    DP-MERF & 113.7 & 80.3 & 66.3 & 62.2 & 203.9 & 27.2 & 176.2 & 81.0 & 278.3 & 60.4 & 106.3 & 81.3 & 106.4 & 62.2 & 214.1 & 29.0 & 147.9 & 81.2 & 251.6 & 58.3 \\
    DP-NTK & 382.1 & 50.0 & 253.1 & 64.4 & 413.0 & 17.0 & 350.4 & 61.2 & 335.5 & 53.1 & 69.2 & 91.3 & 120.5 & 76.3 & 346.9 & 28.2 & 227.8 & 64.2 & 234.5 & 64.1 \\
    DP-Kernel & 33.7 & 94.0 & 63.4 & 68.4 & 184.8 & 26.4 & 140.3 & \underline{83.0} & 254.1 & 68.0 & 38.9 & 93.6 & 74.2 & 70.0 & 161.4 & 25.1 & 128.8 & 83.7 & 217.3 & 68.7  \\
    GS-WGAN & 48.8 & 72.4 & 99.4 & 52.7 & 259.7 & 20.4 & 611.8 & 61.4 & 421.3 & 52.1 & 47.7 & 75.3 & 97.2 & 56.7 & 194.4 & 21.3 & 290.0 & 61.5 &  291.8 & 58.9  \\
    DP-GAN & 57.0 & 92.4 & 74.8 & 71.8 & 187.5 & 26.2 & 112.5 & 77.9 & 132.2 & \underline{83.2} & 30.3 & 92.7 & 76.9 & 70.3 & 138.7 & 30.5 & 31.7 & 89.2 & 66.9 & 79.6  \\
    DPDM & 36.1 & 89.2 & \underline{28.8} & 76.4 & 206.4 & 28.9 & 153.9 & 74.5 & 111.9 & 80.6 & 4.4 & 97.7 & 17.1 & 85.6 & 110.1 & 36.8 & 28.8 & 91.8 & 29.2 &  79.5  \\
    DP-FETA & \underline{13.7} & \underline{95.6} & 31.4 & \underline{81.7} & \underline{139.8} & \underline{35.1} & \underline{60.2} & 82.3 & \underline{52.8} & 77.3 & \underline{3.4} & \underline{98.1} & \underline{13.3} & \underline{87.3} & \underline{95.3} & \underline{43.3} & \underline{24.8} & \underline{94.2} & \underline{27.8} &  \underline{82.9} \\
    \hline
     \rowcolor{gray0} \toolname & \textbf{8.0} & \textbf{97.3}&  \textbf{27.8} & \textbf{83.4}   &  \textbf{120.4} & \textbf{37.9}  & \textbf{48.0} & \textbf{90.0} & \textbf{31.0} & \textbf{84.0} & \textbf{2.5} & \textbf{98.6} & \textbf{11.6}  & \textbf{88.1} & \textbf{69.0} & \textbf{47.0} & \textbf{20.0} & \textbf{95.2} & \textbf{20.3} & \textbf{84.3} \\
    \bottomrule
\end{tabular}
}
\end{table*}

\vspace{0.5mm}
\noindent \textbf{Implementations.} All experiments are implemented with Python 3.8 on a server with 4 NVIDIA GeForce A6000 Ada and 512GB of memory. We aim at conditional generation for these datasets (i.e., each generated image is associated with the class label). Following practical adoption in DPImageBench~\cite{gong2025dpimagebench}, we set DP parameter $\delta = 1/ (N_{\text{priv}} \times \log N_{\text{priv}})$, where $N_{\text{priv}}$ means the number of
samples in training sensitive datasets as presented in~\Cref{tab:datainfo}, and $ \epsilon = \{1, 10\}$. 

Following DPImageBench, the synthesizer sizes are 5.6M parameters for GAN-based methods (DP-MERF, DP-NTK, DP-Kernel, GS-WGAN, DP-GAN) and 3.8M for diffusion-based methods (DPDM, DP-FETA, \toolname). For GAN-based methods, the generator and discriminator sizes are 3.8M and 1.8M parameters, respectively. We recognize that larger synthesizer models may enhance synthetic performance. However, for a fair comparison, this paper adopts the synthesizer size used in prior work. We provide more details of hyper-parameter settings in~\Cref{apsubsec:hyper}.

\vspace{0.5mm}
\noindent \textbf{Investigated Tasks and Datasets.} We perform experiments on five image datasets {\tt MNIST}~\cite{mnist}, {\tt FashionMNIST}~\cite{fmnist} ({\tt F-MNIST}), {\tt CIFAR-10}~\cite{cifar10}, {\tt CelebA}~\cite{celeba}, and {\tt Camelyon}~\cite{camelyon1}. The investigated datasets are prevalently used in previous DP image synthesis methods~\cite{dp-diffusion,dp-feta}. We provide more details of the investigated datasets in~\Cref{apsec:dataset}.

\vspace{0.5mm}
\noindent \textbf{Evaluation Metrics.} We assess the fidelity and utility of synthetic datasets using two widely adopted metrics~\cite{dpdm,dp-feta,gong2025dpimagebench,dpsda,pe3}: Fréchet Inception Distance (FID) and downstream classification accuracy. Our implementation follows the practical framework outlined in DPImageBench~\cite{gong2025dpimagebench}. We generate 60,000 synthetic images for evaluations. Please refer to~\Cref{apsubsec:MetricDetails} for more details.

\section{EMPIRICAL EVALUATIONS}
\label{sec:eval}
This section studies the effectiveness of \toolname{} by answering three Research Questions (RQs) as follows. \textbf{(RQ1)} Does \toolname{} outperform the seven baseline methods across the studied image datasets? \textbf{(RQ2)} How do the frequency features and auxiliary generator benefit warm-up training in \toolname? \textbf{(RQ3)} How do the hyper-parameters of \toolname{} affect the synthesis performance?

\subsection{Performance of Synthetic Datasets (RQ1)}
\label{sec:rq1}

\begin{figure*}[!t]
    \centering
    \setlength{\abovecaptionskip}{0pt}
    \includegraphics[width=1.0\linewidth]{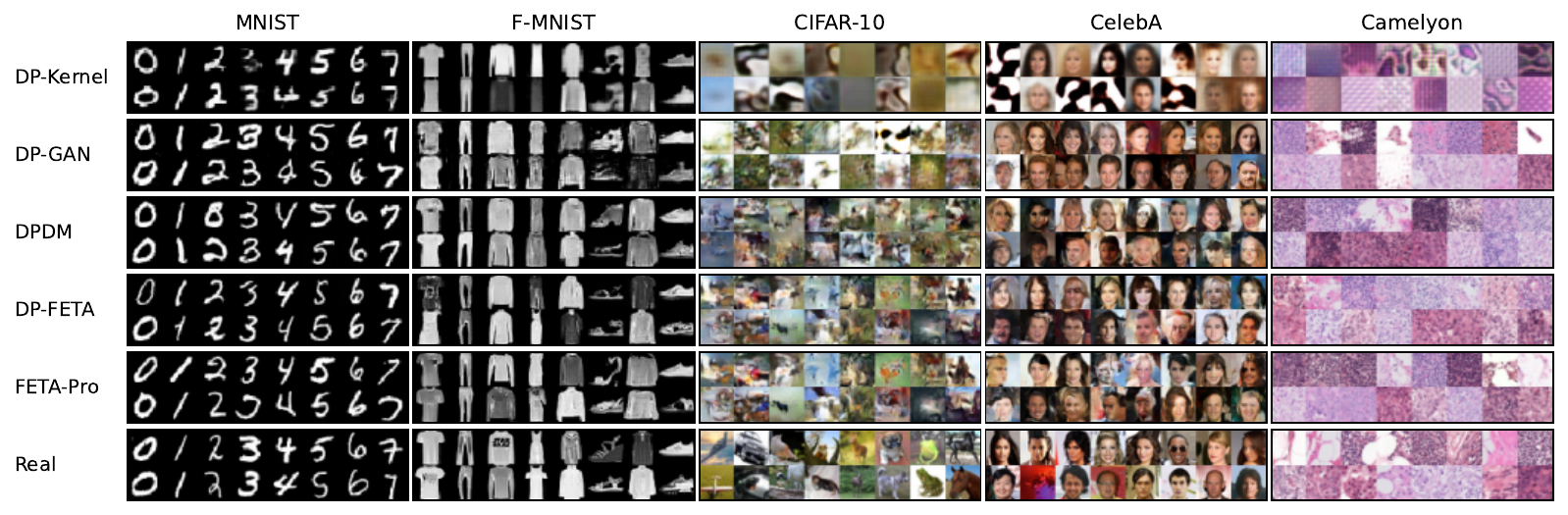}
    \caption{Synthetic Image Examples under $\epsilon = 10$. The last row of images is real image samples from sensitive image datasets. Due to space constraints, we showcase only the top-4 performing baseline methods: DP-Kernel, DP-GAN, DPDM, and DP-FETA.}
    \label{fig:eps10_visual}
\end{figure*}

This RQ investigates the utility and fidelity of synthetic images from \toolname{} and baselines. We compare \toolname{} with seven baselines on five investigated image datasets as described in~\Cref{sec:setup}, under the privacy budget $ \epsilon = \{1,10\}, \delta = 1/ (N_{\text{priv}} \times \log N_{\text{priv}})$. \Cref{fig:eps10_visual} visualizes synthetic image samples generated by \toolname{} and baselines under $\epsilon=10$, alongside real images. 
More details of privacy budget allocation plans are shown in~\Cref{apsubsec:hyper}. 

\Cref{tab:rq1} presents the FID and Acc (\%) of \toolname{} and baselines.  In~\Cref{tab:rq1}, we observe that \toolname{} significantly outperforms the performance of baselines in terms of both utility and fidelity. Specifically, for the {\tt CIFAR-10} and {\tt Camelyon} datasets under $\epsilon = 1$, \toolname{} achieves FID scores of 120.4 and 31.0,
respectively, markedly surpassing the 139.8 and 52.8 scores of DP-FETA. 
Under $\epsilon=10$, \toolname{} attains an FID score of 69.0 on the {\tt CIFAR-10} dataset, outperforming the FID of 95.3 achieved by DP-FETA. Then, the accuracy improves from 43.3\% to 47.0\%. We present more comparisons of fidelity metrics, including Inception Score~\cite{is}, Precision and Recall~\cite{precision&recall}, and Fréchet Leakage Distance~\cite{fld}, for \toolname{} and baseline methods in~\Cref{apsubsec:fidelity}. 

\begin{figure*}[!t]
    \centering
    \setlength{\abovecaptionskip}{0pt}
    \includegraphics[width=1.0\linewidth]{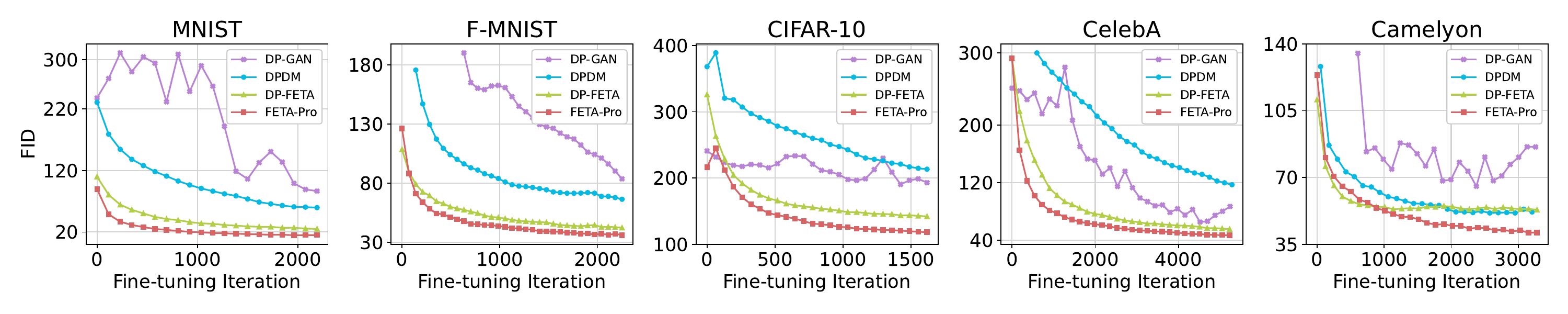}
    \caption{FID of synthetic images during fine-tuning, compared to baseline methods to evaluate convergence under $\epsilon=1$. The FID is computed using 6,000 sampled images (10\% of the final synthetic dataset) relative to the sensitive dataset.}
    \label{fig:fid_curve}
\end{figure*}

We track the FID of synthetic images during DP-SGD fine-tuning to assess convergence against baseline methods. Referring to the previous implementation in DPImageBench~\cite{gong2025dpimagebench}, we sample 6,000 images (10\% of the final synthetic dataset) to compute the intermediate results of FID relative to the sensitive dataset.
\Cref{fig:fid_curve} depicts the FID of synthetic images across various fine-tuning iterations on sensitive images, benchmarked against baseline methods. The figure reveals that synthetic images of \toolname{} converge more rapidly than those of other baselines, enhancing the fidelity of synthetic images. Additionally, compared to DP-GAN, which uses a GAN-based synthesizer, using a diffusion model as the synthesizer offers greater stability. %

\begin{table}[H]
\normalsize
\setlength{\tabcolsep}{3pt}
    \centering
    \renewcommand\arraystretch{1}
    \begin{tabular}{p{1.0\columnwidth}}
    \Xhline{1.0pt}
         \rowcolor{gray0} \noindent \textbf{Answers to RQ1}: Synthetic images produced by \toolname{} show superior fidelity and utility compared to baselines under $\epsilon = 1$. On average, the FID of synthetic images generated by \toolname{} is 25.7\% lower, and the Acc on downstream classification tasks is 4.1\% higher than the SOTA method DP-FETA. The FID of synthetic images of \toolname{} converges more rapidly than that of baselines during training. \\ %
    \Xhline{1.0pt}
    \end{tabular}
\end{table}

\subsection{Benefits of Frequency Features and Auxiliary Generator (RQ2)}
\label{sec:rq2}

We explore the strengths of leveraging the frequency features to warm up the DP image synthesizers. We compare the performance of \toolname{} with three invariants of \toolname{} under $\epsilon=1.0$. 

\begin{itemize}[leftmargin=*]
    \item `\toolnamef' denotes a synthesizer that prioritizes learning \textbf{frequency} knowledge from sensitive datasets before acquiring \textbf{spatial} knowledge.
    \item `\toolnamem' means the synthesizers simultaneously learn the spatial and frequency knowledge.
    \item `\toolnameof' means the synthesizers solely learn the frequency knowledge for warm-up.
\end{itemize}

To thoroughly explore the benefits of warm-up, this RQ examines three perspectives: (1) Is warm-up necessary? (2) Which features are most effective for warm-up? (3) How should the selected features be used for warm-up? 

\begin{figure*}[!t]
    \centering
    \setlength{\abovecaptionskip}{0pt}
    \includegraphics[width=1.0\linewidth]{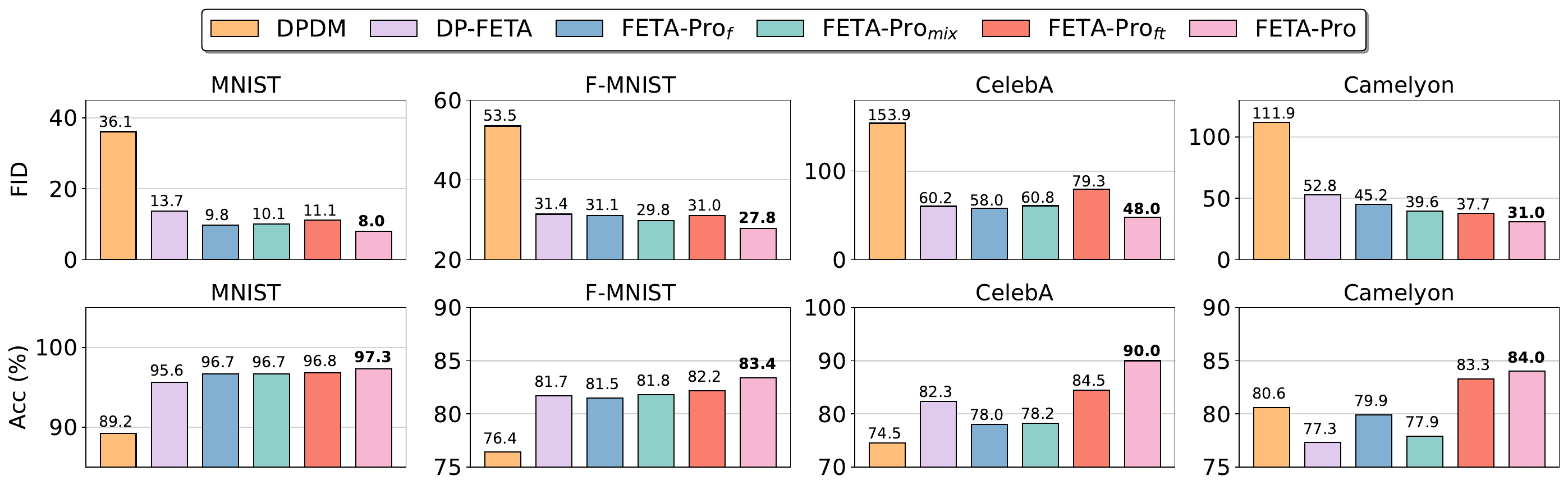}
    \caption{FID (top row) and Acc (bottom row) of \toolname{} and five baselines with $\varepsilon=1$. `DPDM' indicates no warm-up. `DP-FETA' and `\toolnameof' use only spatial and frequency features for warm-up, respectively. `\toolnamem' learns spatial and frequency features simultaneously. `\toolnamef' first learns frequency domain features, then spatial features. `\toolname' is our work, which learns spatial domain features, then frequency features.}
    \label{fig:rq2}
\end{figure*}

\Cref{fig:rq2} presents the FID and Acc of \toolname{} and its variants. For the first question, this figure shows that the training shortcuts enhance synthetic performance. \toolname{} achieves FID reductions of 77.8\% ($=(36.1 - 8.0/36.1)\times 100\%$), 47.9\% ($=(53.5 - 27.8/53.5)\times 100\%$), 68.8\% ($=(153.9 - 48.0/153.9)\times 100\%$),
and 72.3\% ($=(111.9 - 31.0/111.9)\times 100\%$) compared to the no warm up method DPDM.

\begin{table*}[!t]
\renewcommand{\arraystretch}{1.1}
\setlength{\tabcolsep}{5.5pt}
\scriptsize
    \centering
    \caption{Performance of \toolname{} on five sensitive datasets with $\varepsilon=1$. `\toolname-No-Auxiliary' means directly warming up synthesizers on frequency features. `\toolname-DM-Auxiliary' means using diffusion models as the auxiliary generator. `GPU memory' means the peak GPU memory usage across all stages.}
    \label{tab:diffusion_frequnecy_ab}
    \resizebox{1.0\textwidth}{!}{
    \begin{tabular}{l|cccc|cccc|cccc}
    \toprule
    \multirow{2}{*}{\textbf{Dataset}} & \multicolumn{4}{c|}{\toolname} &  \multicolumn{4}{c|}{{\toolname-No-Auxiliary}} &  \multicolumn{4}{c}{{\toolname-DM-Auxiliary}} \\
    \Xcline{2-13}{0.5pt}
     & \centering FID & Acc (\%) & Time & GPU Memory & FID & Acc (\%)  & Time & GPU Memory & FID & Acc (\%)  & Time & GPU Memory \\
    \hline
    {\tt MNIST} & 8.0 & 97.3 & 11.7 H & 73.7 GB &  36.5 & 90.6 & 13.4 H & 74.8 GB & 23.1 & 92.7 & 13.5 H & 73.7 GB\\
    {\tt F-MNIST}  & 27.8 & 83.4 &  11.7 H & 73.7 GB  & 60.5 & 77.2 & 13.4 H & 74.8 GB & 47.3 & 78.4 & 13.5 H & 73.7 GB \\
    {\tt CIFAR-10} & 120.4 & 37.9 &  20.7 H & 96.3 GB  & 237.6 & 29.8 & 27.8 H & 135.8 GB & 230.7 & 29.9 & 27.9 H & 96.3 GB\\
    {\tt CelebA} & 48.0 & 90.0 &  54.4 H & 96.3 GB  & 260.1 & 67.0 & 68.3 H & 135.8 GB & 92.7 & 78.4 & 68.4 H & 96.3 GB\\
    {\tt Camelyon} & 31.0 & 84.0 &  32.9 H & 96.3 GB  & 62.4 & 80.5 & 47.6 H & 135.8 GB & 51.8 & 82.4 & 47.8 H & 96.3 GB\\
    \bottomrule
\end{tabular}
}
\end{table*}

For the second question, this figure highlights that frequency features outperform spatial features for synthesizer warm-up, as analyzed in~\Cref{subsec:theo}. Specifically, \toolnameof{} achieves FID of 9.8, 31.1, 58.0, and 45.2, while DP-FETA, using only spatial features for warm-up, attains reductions of 13.7, 31.4, 60.2, and 52.8 across four datasets.
Additionally, \toolnameof{} outperforms DP-FETA in accuracy, producing higher-quality synthetic images. By combining both feature types, \toolname{} achieves the best synthetic performance among the compared methods. 

For the third question, \toolnamem, which learns frequency and spatial features simultaneously, exhibits poorer synthetic performance than \toolname{} and \toolnamef, which learn these features separately, in most cases. \Cref{fig:rq2} further presents that synthetic images of \toolnamef{} generally outperform DP-FETA in both utility and fidelity,
though they fall short of \toolname{}'s performance. For instance, on the {\tt Camelyon} dataset, \toolnamef{} achieves an FID of 37.7 and an accuracy of 83.3\%, outperforming DP-FETA’s FID of 52.8 and accuracy of 77.3\%, under $\epsilon=1$.
These results verify the analysis introduced in~\Cref{subsec:private_training}. Integrating both spatial and frequency features enhances DP synthesis.
Additionally, frequency features are more intricate than their spatial counterparts, and benefit from the `from easy to hard' learning paradigm. Prioritizing learning simpler features (spatial features) before learning complex features (frequency features) benefits DP synthesis.

To investigate the benefit of auxiliary generators, \Cref{tab:diffusion_frequnecy_ab} compares \toolname{} with two methods mentioned in~\Cref{subsec:moti}. `\toolname-No-Auxiliary' means directly warming up synthesizers on frequency features without leveraging auxiliary generators, as introduced in~\Cref{apsubsec:baselines}.
`\toolname-DM-Auxiliary' means using diffusion models as the auxiliary generator. %
\Cref{tab:diffusion_frequnecy_ab} shows that \toolname-No-Auxiliary exhibits the worst synthetic performance, while auxiliary generators significantly enhance performance. In addition, \toolname{} outperforms \toolname-DM-Auxiliary in terms of both performance and efficiency, showing that GANs are more effective than diffusion models as auxiliary generators.

\begin{table}[H]
\normalsize
\setlength{\tabcolsep}{3pt}
    \centering
    \renewcommand\arraystretch{1}
    \begin{tabular}{p{1.0\columnwidth}}
    \Xhline{1.0pt}
         \rowcolor{gray0} \noindent \textbf{Answers to RQ2}: Using frequency features for synthesizer warm-up proves more effective than using spatial features. Besides, combining both feature types enhances feature complementarity, achieving improvements of 3.6\% in Acc and 10.1 in FID compared to using only frequency features across the studied datasets. Besides, we should prioritize learning spatial features before learning frequency features for better synthetic performance. \\
    \Xhline{1.0pt}
    \end{tabular}
\end{table}

\subsection{The Impact of Hyper-Parameters (RQ3)}
\label{sec:impact_hyper}

This RQ investigates the impact of hyper-parameter settings from two perspectives as follows. We conduct experiments both under the privacy budget $\epsilon=1$ in this section.

\begin{itemize}[leftmargin=*]
    \item \textbf{Privacy allocations.} We evaluate \toolname{} under different privacy allocation plans. As introduced in~\Cref{subsec:privacy_analy}, \toolname{} consumes privacy budget across three stages: (1) extracting spatial features and (2) frequency features from sensitive datasets; and (3) fine-tuning synthesizers on sensitive datasets leveraging DP-SGD. We consider the combination of the noise scales (e.g., $\sigma_t$ and $\sigma_f$) $\{2, 5, 10, 20, 30\}$ and $\{20, 26, 42, 61, 115\}$ for spatial and frequency features. %
    A low level of noise scale means a high level of privacy budget allocation. We ensure that the total privacy budget for these three processes is constrained to $\epsilon=1$. \Cref{apsubsec:privacy_allocation} presents the noise scale of DP-SGD under different privacy allocation strategies.
    \item \textbf{Privacy budget.} We assess the utility and fidelity of synthetic images across six privacy budgets, $\epsilon \in \{0.2, 1.0, 5.0, 10.0, 15.0, 20.0\}$. We adopt the same privacy allocation ratio as described in~\Cref{apsubsec:privacy_allocation}.
\end{itemize}

\begin{figure*}[!t]
    \centering
    \hspace{-0.2cm}
    \subfigure[Acc of synthetic images for {\tt MNIST} (left) and {\tt F-MNIST} (right).]{
        \includegraphics[width=3.43 in]{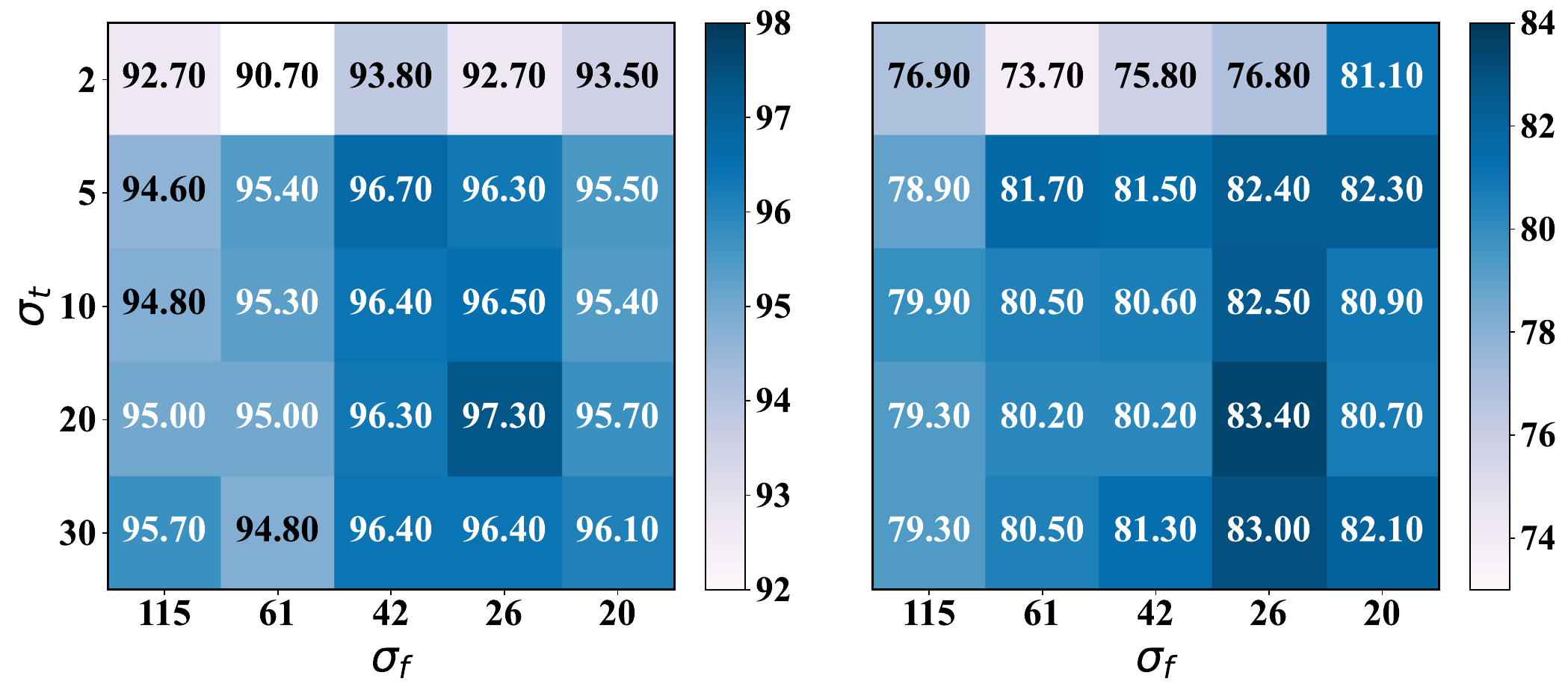}    
    }
    \subfigure[FID of synthetic images {\tt MNIST} (left) and {\tt F-MNIST} (right).]{
        \includegraphics[width=3.43 in]{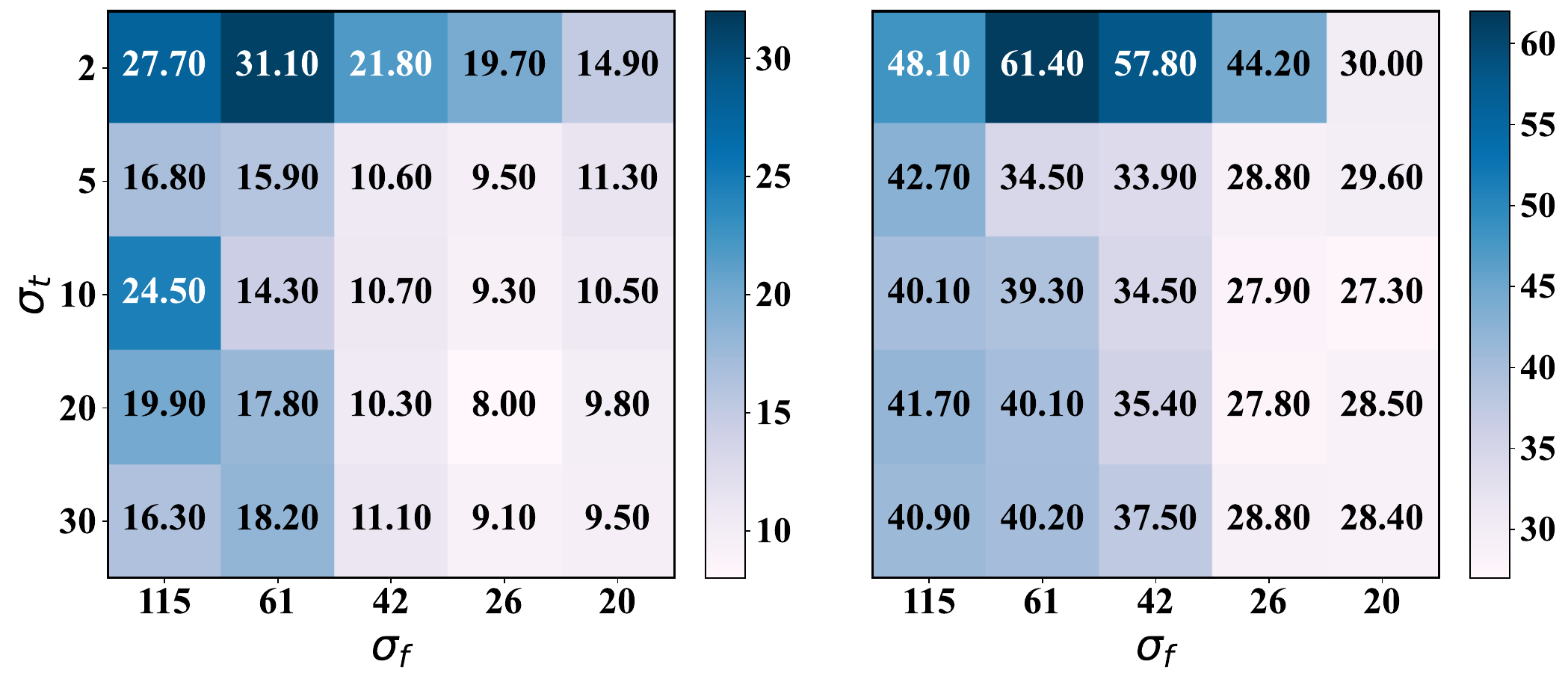}
    }
    \caption{The Acc and FID of synthetic {\tt MNIST} and {\tt F-MNIST} images under $\epsilon = 1$ and $\delta = 1/N \log(N)$. We explore combinations of noise scales $\sigma_t \in \{2, 5, 10, 20, 30\}$ and $\sigma_f \in \{20, 26, 42, 61, 115\}$ (privacy budget from high to low) for spatial and frequency features, ensuring the total privacy budget across all processes (include DP-SGD) is constrained to $\epsilon = 1$.}
    \label{fig:allocation}
\end{figure*}

\Cref{fig:allocation} presents the synthetic performance under different DP privacy budget allocation strategies. We observe that, given the same spatial domain budget allocation, a higher frequency domain budget (low noise scale) allocation generally achieves superior synthetic performance. Specifically, in the {\tt F-MNIST} dataset with a spatial domain noise level of $\sigma_t=2$, the Acc decreases from 81.1 to 76.9 as the frequency domain noise level $\sigma_f$ increases from 20 to 115. Overall, the synthetic performance is sensitive to the value of privacy allocation strategies. In terms of Acc, the best and worst results in~\Cref{fig:allocation} of differences are 6.6\% $(=(97.3-90.7)\times 100\%)$ and 9.7\% $(=(83.4-73.7)\times 100\%)$ for {\tt MNIST} and {\tt F-MNIST}. In terms of FID, the differences are 23.1 $(=(31.1-8.0))$ and 34.1 $(=(61.4-27.3))$. Generally, overly low values of $\sigma_t$ lead to reduced synthetic performance.
For instance, in the {\tt MNIST} dataset, when $(\sigma_t=2, \sigma_f=20)$, \toolname{} achieves an accuracy of 93.5\%, which is noticeably lower than the 97.3\% obtained with $(\sigma_t=20, \sigma_f=26)$. As presented in~\Cref{tab:rdpRatio} of~\Cref{apsubsec:privacy_allocation}, overly low values of $\sigma_t$ indicate that larger privacy budgets are allocated to spatial and domain features, while a smaller budget is assigned to DP-SGD. Generally, privacy budget ratios ordered as `spatial features < frequency features < DP-SGD' can obtain better synthetic performance (referring to the privacy budget ratios in \Cref{tab:rdpRatio} of \Cref{apsubsec:privacy_allocation}). We provide the relatively optimal combinations of $\sigma_f$ and $\sigma_t$ values in~\Cref{tab:hyper-parameter} of~\Cref{apsubsec:hyper} for different sensitive datasets.

\begin{figure}[!t]
    \centering
    \setlength{\abovecaptionskip}{0pt}
    \includegraphics[width=0.49\textwidth]{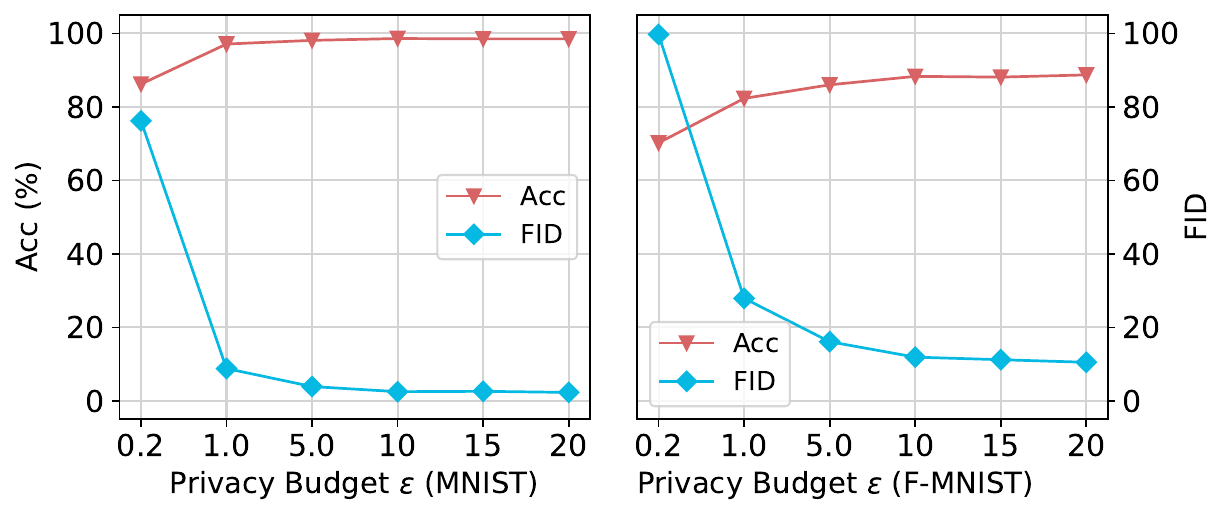}
    \caption{The Acc and FID of synthetic images generated by \toolname{} for {\tt MNIST} and {\tt F-MNIST} under privacy budgets $\epsilon = \{0.2, 1.0, 5.0, 10, 15, 20\}$ and a fixed $\delta = 1/N \log(N)$. }
    \label{fig:change_eps}
\end{figure}

In terms of Acc, \toolname{} achieves optimal results of 97.3\% and 83.4\% for the {\tt MNIST} and {\tt F-MNIST} datasets, respectively, at $(\sigma_t=20, \sigma_f=26)$. In terms of FID, \toolname{} achieves optimal results of 8.0 and 27.3, at $(\sigma_t=20, \sigma_f=26)$ with a spatial-to-frequency budget cost ratio of  $0.3/2.92$ and $(\sigma_t=10, \sigma_f=20)$ with a ratio of $1.22 / 4.99$, as presented in \Cref{tab:rdpRatio} of \Cref{apsubsec:privacy_allocation}. These results indicate that superior performance is achieved by allocating a greater privacy budget to frequency features compared to spatial features.

\Cref{fig:change_eps} presents the Acc and FID scores under different privacy budgets. We observe that a higher privacy budget achieves improved synthetic performance. For the {\tt MNIST} dataset, when $\epsilon\geq 10$, the Acc approaches 100\%, and the FID nears 0, indicating that the synthetic images closely resemble the original ones. Similarly, for {\tt F-MNIST}, performance plateaus at $\epsilon \geq 10$, suggesting that $\epsilon = 10$ balances synthetic performance with privacy preservation in practice.

\begin{table}[H]
\normalsize
\setlength{\tabcolsep}{3pt}
    \centering
    \renewcommand\arraystretch{1}
    \begin{tabular}{p{1.0\columnwidth}}
    \Xhline{1.0pt}
         \rowcolor{gray0} \noindent \textbf{Answers to RQ3}: Allocation strategies highly affect synthetic image quality. Overall, privacy budget ratios in \toolname{} ordered as `spatial features < frequency features < DP-SGD' can obtain better performance. For both {\tt MNIST} and {\tt F-MNIST}, performance stabilizes at $\epsilon\geq 10$, indicating that $\epsilon=10$ balances synthetic quality with privacy preservation in practice. \\
    \Xhline{1.0pt}
    \end{tabular}
\end{table}

\section{Discussions}
\label{sec:dis}

This section discusses (1) the impact of the DP on the performance of \toolname, (2) how \toolname{} performs when using public images, and (3) the computational resources requirements. \Cref{subsec:limit} introduces the limitations of \toolname.

\begin{table}[!t]
\renewcommand{\arraystretch}{1.1}
\setlength{\tabcolsep}{5.5pt}
\small
    \centering
    \caption{FID and Acc (\%) of \toolname{} on four sensitive datasets with $\varepsilon=\{10,\infty\}$. `No DP' means the classifier is trained directly on the sensitive dataset.}
    \label{tab:no_dp}
    \resizebox{0.49\textwidth}{!}{
    \begin{tabular}{l|cc|cc|cc|cc}
    \toprule
    \multirow{2}{*}{\textbf{Method}} & \multicolumn{2}{c|}{{\tt MNIST}} & \multicolumn{2}{c|}{{\tt F-MNIST}}  &  \multicolumn{2}{c|}{{\tt CelebA}} & \multicolumn{2}{c}{{\tt Camelyon}} \\
    \Xcline{2-9}{0.5pt}
     & \centering FID & Acc & FID & Acc & FID & Acc & FID & Acc \\
    \hline
    No DP & - & 99.7 & - & 94.5 & - & 97.7 & - & 87.7 \\
    Ours $(\epsilon=10)$  &  2.5 & 98.6 & 11.6  & 88.1 & 20.0 & 95.2 & 20.3 & 84.3 \\
    Ours $(\epsilon=\infty)$ & 1.1 & 99.3 & 5.7 & 91.0 & 7.0 & 95.3 & 6.6 & 86.3 \\
    \bottomrule
\end{tabular}
}
\end{table}

\subsection{\toolname{} without Privacy Protection}
\label{subsec:without_protection}

This experiment evaluates the impact of the DP on the synthetic performance of \toolname{}. We compare \toolname{} with three baseline methods: (1) `$\epsilon=\infty$' means that diffusion models are trained using the \toolname{} framework without introducing Gaussian noise during training. (2) `No DP': The classifier is trained directly on the sensitive dataset for the downstream classification task. Note that the FID evaluates the quality of images generated by generative models; thus, sensitive datasets do not have an FID score. %

\Cref{tab:no_dp} shows that, under a privacy budget of $\epsilon=10$, \toolname{} achieves average accuracy (Acc) reductions of only 0.7\% $(=(99.3-98.6)\times 100\%)$, 2.9\% $(=(91.0-88.1)\times 100\%)$,
0.1\% $(=(95.3-95.2)\times 100\%)$, and 2.0\% $(=(86.3-84.3)\times 100\%)$ across four sensitive datasets compared to the non-private setting ($\epsilon=\infty$). As highlighted in previous work~\cite{hod2024differentially}, it is common to use $\epsilon \leq 10$ in differential practical synthesis tasks. Therefore, \toolname{} generates highly useful synthetic images while protecting sensitive data. Besides, the FID decreases by 1.4 ($=(2.5-1.1))$, 5.9 $(=(11.6-5.7))$, 13.0 $(=(20.0-7.0))$, and 13.7 $(=(20.3-6.6))$. Compared to directly using sensitive datasets for downstream tasks (i.e., `No DP' in \Cref{tab:no_dp}), \toolname{} exhibits an average Acc reduction of 3.4\%. These results indicate the need for further improvements to \toolname{}.

\subsection{\toolname{} Leveraging Public Images}
\label{subsec:leverage_public}

This section evaluates whether pre-training with public datasets enhances the synthetic performance of \toolname{} and compares \toolname{} with state-of-the-art methods using public images, as detailed below. Referring to DPImageBench~\cite{gong2025dpimagebench}, we leverage {\tt ImageNet}~\cite{imagenet} as the pre-training dataset.

\begin{itemize}[leftmargin=*]
\item \textbf{PDP-Diffusion~\cite{dp-diffusion}:} PDP-Diffusion adopts the `public pre-training + private fine-tuning' paradigm. It incorporates large batch sizes to improve the stability and accelerate the convergence of diffusion model training under DP-SGD. Moreover, leveraging public datasets for pre-training allows the synthesizer to benefit from a broader knowledge base.

\begin{table}[!t]
\renewcommand{\arraystretch}{1.1}
\setlength{\tabcolsep}{5.5pt}
\small
    \centering
    \caption{FID and Acc of \toolname{} and baselines which use public images on four image datasets with $\varepsilon=1$. The best performance in each column is highlighted using a bold font.}
    \label{tab:public}
    \resizebox{0.49\textwidth}{!}{
    \begin{tabular}{l|rc|rc|rc|rc}
    \toprule
    \multirow{2}{*}{\textbf{Method} ($\varepsilon=1$)} & \multicolumn{2}{c|}{{\tt MNIST}} & \multicolumn{2}{c|}{{\tt F-MNIST}}
    & \multicolumn{2}{c|}{{\tt CelebA}} & \multicolumn{2}{c}{{\tt Camelyon}}\\
    \cline{2-9}
     & \centering FID & Acc & FID & Acc & FID & Acc & FID & Acc\\
    \hline
    PDP-Diffusion & 8.9 & 94.5 & 16.6 & 79.2 & 17.2 & 89.4 & \textbf{15.0} & \textbf{85.2} \\
    PrivImage & \textbf{7.6} & 94.0 & \textbf{16.1} & 79.9 & \textbf{12.3} & \textbf{90.8} & 15.2 & 82.8 \\
    PE & 48.0 & 27.9 & 48.8 & 47.9 & 23.0 & 70.5 & 71.5 & 63.3 \\
    \toolname{}  & 8.0 & \textbf{97.3} &  27.8 & \textbf{83.4}  & 48.0 & 90.0 &  31.0 & 84.0  \\
    \hline
    PDP-Diffusion-Pro & 10.9 & 95.3 & 23.4 & 81.3 & 25.2 & 90.1 & 17.2 & 84.1\\
    PrivImage-Pro & 9.9 & 94.7 & 25.4 & 78.2 & 23.1 & 90.2 & 15.9 & 84.4 \\
    \bottomrule
\end{tabular}
}
\end{table}

\item \textbf{PrivImage~\cite{li2023PrivImage}:} Unlike PDP-Diffusion, which uses entire public datasets for pre-training, PrivImage selectively queries the semantic distribution of sensitive data to identify a subset of public data, narrowing the distribution gap between public and sensitive datasets and enhancing the efficiency of pre-training~\cite{li2023PrivImage}.
\item \textbf{Private Evolution (PE)~\cite{dpsda}:} PE progressively guides the pretrained models (either local or blackbox models behind APIs) to generate synthetic images resembling a sensitive dataset, eliminating the need for training or fine-tuning.
\end{itemize}

\Cref{tab:public} compares the FID and accuracy of \toolname{} and baseline methods using public images across four image datasets with $\epsilon = 1$. The `PDP-Diffusion-Pro' and `PrivImage-Pro' variants incorporate the warm-up method from \toolname{} into PDP-Diffusion~\cite{dp-diffusion} and PrivImage~\cite{li2023PrivImage}, respectively. These two variant methods follow the `public pre-training + private warm-up + private fine-tuning' paradigm. \toolname{} is our proposed method in this work.

\begin{table}[!t]
\setlength{\tabcolsep}{4.5pt}
\small
\centering
\caption{GPU memory usage and runtime analysis of studied methods for the {\tt CIFAR-10}. ‘GPU memory’ means the peak GPU memory usage across all stages.}
\setlength{\tabcolsep}{2.6mm}{
\resizebox{0.47\textwidth}{!}{
\begin{tabular}{l|l|rrc}
\toprule
\textbf{Algorithm} & \textbf{Stage} & \textbf{Memory} & \textbf{Runtime} & \textbf{GPU Memory} \\
\midrule
\multirow{3}{*}{DP-MERF} & Warm-up & 0 GB & 0 H & \multirow{3}{*}{18.6 GB} \\
                         & Fine-tune & 3.5 GB & 0.02 H & \\
                         & Synthesis & 18.6 GB & 0.03 H & \\
\hline
\multirow{3}{*}{DP-NTK}  & Warm-up & 0 GB & 0 H & \multirow{3}{*}{21.3 GB} \\
                         & Fine-tune & 4.8 GB & 7.75 H & \\
                         & Synthesis & 19.1 GB & 0.05 H & \\
\hline
\multirow{3}{*}{DP-Kernel} & Warm-up & 0 GB & 0 H & \multirow{3}{*}{22.5 GB} \\
                           & Fine-tune & 8.8 GB & 3.6 H & \\
                           & Synthesis & 19.0 GB & 0.05 H & \\
\hline
\multirow{3}{*}{GS-WGAN}   & Warm-up & 21.6 GB & 18.3 H & \multirow{3}{*}{21.6 GB} \\
                           & Fine-tune & 21.6 GB & 0.6 H & \\
                           & Synthesis & 20.0 GB & 0.01 H & \\
\hline
\multirow{3}{*}{DP-GAN}    & Warm-up & 0 GB & 0 H & \multirow{3}{*}{46.1 GB} \\
                           & Fine-tune & 46.1 GB & 3.25 H & \\
                           & Synthesis & 20.1 GB & 0.03 H & \\
\hline
\multirow{3}{*}{DPDM}      & Warm-up & 0 GB & 0 H & \multirow{3}{*}{96.3 GB}  \\
                           & Fine-tune & 96.3 GB & 20.5 H &  \\
                           & Synthesis & 32.1 GB & 0.17 H &  \\
\hline
\multirow{3}{*}{DP-FETA}      & Warm-up & 3.4 GB & 0.04 H &  \multirow{3}{*}{96.3 GB}\\
                           & Fine-tune & 96.3 GB & 20.5 H \\
                           & Synthesis & 32.1 GB & 0.17 H &  \\
\hline
\multirow{3}{*}{\toolname}      & Warm-up & 3.4 GB &  0.1 H &  \multirow{3}{*}{96.3 GB}\\
                           & Fine-tune & 96.3 GB &  20.5 H \\
                           & Synthesis & 32.1 GB & 0.17 H &  \\
\bottomrule
\end{tabular}}}
\label{tab:computationalResource}
\end{table}

Comparing the results in rows four through six of \Cref{tab:public}, we observe that public pre-training enhances synthetic performance on \texttt{CelebA} and \texttt{Camelyon} datasets, but reduces performance on \texttt{MNIST} and accuracy on \texttt{F-MNIST}.
Overall, warm-ups do not always benefit the algorithms. Specifically for the \texttt{MNIST} dataset, \toolname{} achieves an FID of 8.0 and an accuracy of 97.3\%. This performance surpasses methods that utilize public resources, such as PDP-Diffusion (FID: 8.9, Acc: 94.5\%) and PE (FID: 48.0, Acc: 27.9\%). Although the FID achieved by \toolname{} is 0.4 lower than PrivImage (8.0 vs. 7.6), its Acc is 3.3 percentage points higher (97.3\% vs. 94.0\%).
Pre-training on public datasets may disrupt \toolname{}'s warm-up process, leading to degraded performance. Specifically, the FID increases from 8.0 to 10.9, and the accuracy drops from 97.3\% to 95.3\%
compared to PDP-Diffusion-Pro.

For {\tt CelebA} and {\tt Camelyon}, pre-training enhances synthetic performance so that PDP-Diffusion-Pro and PrivImage-Pro result in lower FID and higher accuracy compared to \toolname{}. While PDP-Diffusion-Pro and PrivImage-Pro show improved synthetic performance compared to FETA-Pro, their comparison against PDP-Diffusion and PrivImage achieves mixed results. We can observe that in {\tt CelebA}, PrivImage-Pro achieves better Acc (90.2\%) than PDP-Diffusion (89.4\%). However, PDP-Diffusion-Pro and PrivImage-Pro achieve higher (and thus worse) FID scores than both PDP-Diffusion and PrivImage. Their best accuracies (PrivImage-Pro: 90.2\% for {\tt CelebA}, 84.4\% for {\tt Camelyon}) still do not surpass the peak performance observed in PDP-Diffusion (85.2\% for {\tt Camelyon}) and PrivImage (90.8\% for {\tt CelebA}).
Overall, we advise against combining pre-training with the warm-up, as these operations may interact detrimentally, leading to suboptimal results.

\subsection{Comparison of Computational Resources}

This section investigates the computational resource usage of various methods. The parameter sizes of DP image synthesizers and the servers used for implementation are detailed in~\Cref{sec:setup}. \Cref{tab:computationalResource} presents the GPU memory and runtime usage of baselines and \toolname. Compared to the SOTA DP image synthetic method DP-FETA, which only uses spatial features to warm-up, \toolname{} only introduces an average of 0.3\% additional training-time cost (i.e., 0.06 H) for frequency features warming up while bringing increases in fidelity and utility metrics as presented in~\Cref{tab:rq1}. Additionally, \toolname{} incurs no extra memory overhead compared to DP-FETA.

\section{Related Work}
\label{sec:related}
We discuss related work briefly here, and we provide a more comprehensive discussion in Appendix~\ref{suppsec:related}. We discuss previous DP image synthesis methods based on whether the methods leverage public resources or not. 

\vspace{1mm}
\noindent \textbf{Without using Public Resources.} These methods train the synthesizer solely using the sensitive image dataset, without relying on external public resources~\cite{dp-merf,dp-feta,pearl,dp-ntk}. A group of methods proposes adding noise to the high-level feature for DP, e.g., random Fourier features~\cite{dp-mepf,pearl}, empirical neural tangent kernels~\cite{dp-ntk}, of sensitive images, and training a synthesizer to ensure synthetic images approach noisy features. 

Although the aforementioned methods have low computational requirements, they present suboptimal synthetic performance on complex datasets, such as {\tt CelebA}~\cite{celeba}. Abadi et al.~\cite{dpsgd} introduce DP-SGD, which adds noise to training gradients to achieve DP. Various works use DP-SGD to train deep generative models on sensitive datasets for addressing DP synthesis of complex images~\cite{dpgan,g-pate,dpdm,dpldm,dp-diffusion,li2023PrivImage,dplora}. Among these approaches, utilizing diffusion models as synthesizers achieves SOTA synthetic performance~\cite{dpdm,dp-diffusion,li2023PrivImage,dpldm}.

\vspace{1mm}
\noindent \textbf{Using Public Dataset.} Several methods suggest using a publicly available dataset to initially train a data synthesizer, followed by fine-tuning models with sensitive images using DP-SGD, to improve synthetic performance~\cite{dp-mepf,chen2022dpgen,dp-diffusion,dplora,dpldm,yin2022practical,lin2020using}. For example, Ghalebikesabi et al.~\cite{dp-diffusion} apply this pre-training and fine-tuning method using diffusion models. Li et al.~\cite{li2023PrivImage} propose PrivImage, which enhances synthetic data utility by selecting a subset of public data that aligns distributionally with the sensitive dataset. This paradigm represents the predominant framework for DP image synthesis.

\vspace{1mm}
\noindent \textbf{Using Pretrained Models.} 
Fine-tuning with DP-SGD remains computationally intensive and is time-consuming~\cite{gong2025dpimagebench}. Lin et al.~\cite{dpsda} propose a fine-tuning-free method, Private Evolution (PE). This approach iteratively directs pre-trained models to produce synthetic images that closely align with sensitive images in the feature space. 
However, when the distribution of synthetic data generated by the pre-trained models diverges from that of the sensitive data, PE’s performance underperforms compared to fine-tuning-based methods~\cite{pe3,gong2025dpimagebench}. To address this limitation, Lin et al.~\cite{pe3} introduce Sim-PE, an extension of PE that uses powerful non-neural-network simulators (e.g., avatar generator~\cite{pythonavatar}, computer graphics tools) instead of pre-trained models to fit sensitive data under DP.

\section{Conclusions}
\label{sec:con}

This paper investigates the incorporation of different training shortcuts to augment DP-SGD. \toolname{} integrates frequency features, which robustly capture global structures and textures, as the training shortcut from two key perspectives: (1) they complement spatial features, and (2) their complexity, falling between that of spatial features and full images, enables a fine-grained curriculum framework. This allows \toolname{} to learn sequentially from `spatial features $\to$ frequency features $\to$ images.' When combining these two types of shortcuts, we should address the challenges of training discrepancy between spatial and frequency features. To do so, instead of training one model on multiple features or objectives, we leverage the pipeline generation property of generative models. This allows us to use a more flexible design by having multiple models work on different features. 
Specifically, \toolname{} introduces an \textit{auxiliary generator} to produce images from noisy frequency features. Then, another model is trained with these generated images, alongside spatial features and DP-SGD. Thus, \toolname{} represents both feature types in a unified manner.  We conduct experiments across five sensitive image datasets, showing that \toolname{} achieves better synthetic images compared to the state-of-the-art methods. By integrating the strengths of multiple synthesizers, \toolname{} surpasses traditional single-synthesizer approaches, offering new insights for future research in DP image synthesis.

\subsection*{Ethical considerations.} 

\rev{We structure the ethical considerations discussion by linking our \textit{stakeholder analysis} to the impacts generated during two distinct phases: the \textit{research process} (data handling) and the \textit{publication of results} (deployment and application). We then detail \textit{mitigations} (specifying which stakeholder group each measure protects) and conclude with the \textit{justification} for conducting this research.

\vspace{1mm}
\noindent \textbf{Stakeholder Analysis and Process Impact.} \toolname{} involves three primary stakeholder groups as follows. (1) \textit{Data Subjects}: These are the individuals represented in sensitive datasets (e.g., medical scans, faces). During the research process, these subjects rely on requirements to confidentiality protocols to ensure their sensitive visual data is not exposed. (2) \textit{Data Owners (Institutions)}: Entities such as hospitals and research institutions hold these sensitive datasets. Throughout the research process, these entities face challenges due to privacy regulations and legal restrictions that hinder collaboration. (3) \textit{Researchers and Practitioners:} This group includes the ML and security community. They rely on the process to produce methodologically sound and reproducible DP.

\vspace{1mm}
\noindent \textbf{Impact of the Research.} The publication of \toolname{} has both positive and negative impacts, directly affecting the stakeholders identified above.

\vspace{1mm}
\noindent \textit{Positive Impacts.} (1) Facilitating Data Sharing (Impact on Data Owners \& Researchers): \toolname{} improves the fidelity of DP synthetic images. We provide data owners with a tool to share data utility without violating privacy regulations. This directly benefits Researchers by enabling access to previously inaccessible data for training ML models. (2) Reproducibility and Open Science (Impact on Practitioners): We have open-sourced our code and provided detailed explanations to ensure reproducibility. This allows the practitioner community to responsibly assess, audit, and improve upon our methods, ensuring that future applications in real-world scenarios are built on a transparent foundation.

\vspace{1mm}
\noindent \textit{Negative Impacts.} (1) Bias Propagation (Impact on Data Subjects): We acknowledge that DP synthetic images may inherit or even amplify biases present in the original data (e.g., gender or racial biases). If practitioners train models on synthetic images, the resulting systems may exhibit reduced fairness. This negatively impacts data subjects, who may be affected by decisions made by models with ingrained biases. (2) Potential for Misuse (Impact on Data Subjects \& Society): The publication of high-fidelity synthesis methods carries the risk of misuse. Malicious actors could use these methods to generate realistic fake content (forging identities or fraud) or to train invasive surveillance systems.

\vspace{1mm}
\noindent \textbf{Mitigation.} To address the risks outlined above, we have implemented mitigation strategies. We clarify below how each technical measure serves to protect a specific stakeholder.

\vspace{1mm}
\noindent \textit{Methodological Mitigations (Implemented).} (1) Protecting Data Subjects via Data Curation: To safeguard data subjects from the amplification of external biases, we avoid using public datasets for pretraining~\cite{position}. Public datasets may contain uncontrolled biases (e.g., category imbalances) or inadvertent sensitive content. By restricting pretraining, we minimize the risk that our synthetic images inherit prejudices. (2) Ensuring Reliability for Practitioners via Validation: We incorporate rigorous validation steps to detect potential inaccuracies in synthetic images. While this ensures technical correctness, its primary ethical function is to protect practitioners from drawing false conclusions. By preventing the downstream propagation of flawed data, we ensure that models built on our synthesis do not result in erroneous decisions. (3) Empowering the Community via Transparency: We release our code. This empowers the research community to audit our privacy claims and allows institutions to verify the safety before applying it to their own sensitive data.

\vspace{1mm}
\noindent \textit{Recommended Future Deployment Measures.} Beyond our methodological choices, we advocate for the following measures to protect stakeholders in reality. (1) Fairness Audits for Data Subjects: To further protect data subjects from algorithmic discrimination, we recommend that practitioners integrate fairness constraints~\cite{christopher2024constrained}. The evaluation should be expanded to include comprehensive fairness metrics~\cite{liu2025can} to ensure that the synthetic data does not marginalize specific demographic groups. (2) Safeguards Against Misuse for Society: To address the societal risk of deepfakes, we suggest the implementation of DP watermarking~\cite{an2024waves}. Inserting indelible watermarks into synthetic datasets allows tracing the origin of the data, serving as a deterrent against malicious actors.

\vspace{1mm}
\noindent \noindent \textbf{Justification for Research.} Finally, we posit that the benefits of this research outweigh the potential risks, particularly given the mitigations in place. This work addresses a critical challenge in privacy-preserving machine learning: balancing utility and privacy without relying on risky public datasets. Conducting this research is essential for the following reasons: (1) \toolname{} presents a novel integration of spatial and frequency-domain features, contributing new insights into the design for privacy-preserving generative models; (2) We share the code of our method to promote transparency and community-driven improvements, reducing the risk of opaque implementations. Our repository also serves as a step towards identifying potential risks in DP image synthesis.}

\subsection*{Open Science}

We release the replication package on the anonymous link,\footnote{\label{link}\url{https://github.com/2019ChenGong/Feta-Pro}}. In the {\tt README.md} files of the repository, we provide a clear, point-by-point explanation that maps each table and figure to the corresponding code needed to generate its results. The  DOI for the artifacts is on Zenodo.\footnote{\url{https://doi.org/10.5281/zenodo.17836341}}

We provide a detailed description of how our repository matches each table and figure in our paper as follows.

\vspace{1mm}
\noindent \textbf{Environment:} We provide a straightforward environment setup. The user only needs to run ``bash install.sh'' to install our environment in minutes, supporting availability and functionality assessments.

\vspace{1mm}
\noindent \textbf{Data Preparation:} This artifact provides a concrete data preparation process. All datasets used are publicly available and widely used in the community to avoid ethical concerns. The user should run ``bash data\_preparation.sh'' to prepare all the studied datasets. Loading all datasets is time‑consuming. For quick evaluation, we recommend loading only {\tt MNIST} and {\tt Fashion‑MNIST}, using ``bash data\_preparation\_quick.sh''. 

\noindent \textbf{Main Contributions:} Our artifices provides the reproducibility of tables and figures presented in the main body as follows.
\vspace{-3.5mm}
\begin{itemize}[leftmargin=*]
    \item \texttt{\Cref{tab:rq1}}: In RQ1, FID and Acc (\%) of \toolname{} and seven baselines on {\tt MNIST}, {\tt F-MNIST}, {\tt CIFAR-10}, {\tt CelebA} and {\tt Camelyon} with $\varepsilon=\{1,10\}$.
    \vspace{-2.0mm}
    \item \texttt{\Cref{tab:diffusion_frequnecy_ab}}: In RQ2, performance of \toolname{} on five sensitive datasets with $\varepsilon=1$, compared to `\toolname-No-Auxiliary' and `\toolname-DM-Auxiliary' (the variants of \toolname{}).
    \vspace{-2.0mm}
    \item \texttt{\Cref{tab:no_dp}}: In Discussion, FID and Acc (\%) of \toolname{} on four sensitive datasets with $\varepsilon=\{10,\infty\}$. `No DP' means the classifier is trained directly on the sensitive dataset.
    \vspace{-2.0mm}
    \item \texttt{\Cref{tab:public}}: In Discussion, FID and Acc of \toolname{} and baselines (PDP-Diffusion and PrivImage) which use public images on four image datasets with $\varepsilon=1$.
    \vspace{-2.0mm}
    \item \texttt{\Cref{tab:add_metrics}}: In \Cref{apsubsec:fidelity}, The IS, Precision, Recall, and FLD of \toolname{} and seven baselines on five studied datasets with $\varepsilon=\{1,10\}$.
    \vspace{-2.0mm}
    \item \texttt{\Cref{tab:prv_rdp}}: In \Cref{apsubsec:privacy_allocation}, FID and Acc (\%) of \toolname{} on five sensitive datasets with $\varepsilon=10$ using RDP and PRV privacy budget accounting methods.
    \vspace{-2.0mm}
    \item \texttt{\Cref{tab:rdpRatio}}: In \Cref{apsubsec:privacy_allocation}, RDP cost ratios (\%) of spatial features / frequency features / DP-SGD in \toolname{} using various privacy allocation strategies, under the $\epsilon=1.0$.
    \vspace{-2.0mm}
    \item \texttt{\Cref{fig:eps10_visual}}: In RQ1, synthetic image examples under $\epsilon = 10$.
    \vspace{-5.0mm}
    \item \texttt{\Cref{fig:fid_curve}}: In RQ1, FID of synthetic images during fine-tuning, compared to baseline methods to evaluate convergence under $\epsilon=1$.
    \vspace{-2.0mm}
    \item \texttt{\Cref{fig:rq2}}: In RQ2, FID and Acc of \toolname{} and five baselines with $epsilon=1$. `DPDM' indicates no warm-up. `DP-FETA' and `\toolnameof' use only spatial and frequency features for warm-up, respectively. `\toolnamem' learns spatial and frequency features simultaneously. `\toolnamef' first learns frequency domain features, then spatial features. `\toolname' is our work.
    \vspace{-2.0mm}
    \item \texttt{\Cref{fig:allocation}}: In RQ3, Acc and FID of synthetic {\tt MNIST} and {\tt F-MNIST} images under $\epsilon = 1$ and different privacy budget allocation plans.
    \vspace{-2.0mm}
    \item \texttt{\Cref{fig:change_eps}}: In RQ3, The Acc and FID of synthetic images generated by \toolname{} for {\tt MNIST} and {\tt F-MNIST} under privacy budgets $\epsilon = \{0.2, 1.0, 5.0, 10, 15, 20\}$.
\end{itemize}
\vspace{-2.0mm}

\bibliographystyle{ieeetr}
\bibliography{bib}

\appendix

\setcounter{section}{0}
\setcounter{equation}{0}
\renewcommand\thesection{\Alph{section}}

\section{Details of R\'{e}nyi DP}
\label{app:supp_dp}

This section outlines the computation of privacy costs using R\'{e}nyi DP (RDP)~\cite{sgm}, which offers a robust framework for tracking privacy loss in DP mechanisms.

\begin{definition}[Sub-sampled Gaussian Mechanism (SGM~\cite{sgm})]
    Let $f:{D_s} \subseteq D \to {\mathbb{R}^d}$ be a function with sensitivity ${\Delta _f} = \max_{D\simeq D'}{\left\| {f\left(D \right) - f\left({D'} \right)} \right\|_2}$.  Parameterized with a sampling rate $q \in \left( {0,1} \right]$ and noise standard deviation $\sigma>0$, the SGM $\mathcal{Q}$ is defined as,
\[
    \mathcal{Q}_{f,q,\sigma }\left( D \right) \buildrel \Delta \over = f\left( S \right) + \mathcal{N} \left( {0,{\sigma ^2}\Delta_f^2{\mathbb{I}}} \right)
    \nonumber\]
\end{definition}

\noindent where $S =$ \{${x\left| x \in D \right.}$ selected independently with probability $q$\} and $f\left( \varnothing  \right) = 0$. The privacy loss of SGM can be tracked through R\'{e}nyi DP~\cite{sgm}, as elaborated in~\Cref{def:rdp}. RDP can quantify the privacy loss of SGM accurately, as introduced in~\Cref{eq:rdp_gamma}. 

\begin{definition}[\textit{R\'{e}nyi DP}~\cite{sgm}]
\label{def:rdp}
The R\'{e}nyi divergence between two probability distributions $Y$ and $N$ is defined as, $D_\alpha(Y \| N) = \frac{1}{\alpha - 1} \ln \mathbb{E}_{x \sim N} \left[ \left( \frac{Y(x)}{N(x)} \right)^\alpha \right],$
where $\alpha > 1$ is a real number. A randomized mechanism $\mathcal{A}$ satisfies $(\alpha, \gamma)$-RDP if, for any neighboring datasets $D$, $D'$, and algorithm $\mathcal{Q}$, it holds that, $D_\alpha(\mathcal{Q}(D) \| \mathcal{Q}(D')) \leq \gamma$.
\end{definition}

\begin{table*}[!t]
    \centering
    \caption{Hyper-parameter settings of \toolname. `LR' denotes Learning Rate.}
    \label{tab:hyper-parameter}
    \resizebox{1.0\textwidth}{!}{
    \renewcommand{\arraystretch}{1.1}
    \begin{tabular}{l|ccccc|ccccc}
    \toprule
    \multirow{2}{*}{\textbf{Hyper-parameter}}& \multicolumn{5}{c|}{$\epsilon=1.0$} & \multicolumn{5}{c}{$\epsilon=10.0$}\\
    \Xcline{2-11}{0.5pt}
    & {\tt MNIST}& {\tt F-MNIST}&
    {\tt CIFAR-10}&{\tt CelebA} & {\tt Camelyon}& {\tt MNIST}& {\tt F-MNIST}&{\tt CIFAR-10}& {\tt CelebA} & {\tt Camelyon}\\
    \hline
    Noise scale $\sigma_f$ & 26.6 & 26.6 & 32.6 & 25.0 & 49.6 & 7.4 & 7.4 & 8.2 & 8.2 & 8.4\\
    Noise scale $\sigma_t$ & 20 & 20 & 15 & 15 & 10 & 5 & 5 & 5 & 50 & 5 \\
    Spatial domain Epoch & 1,000 & 1,000 & 1,000 & 1,000 & 1,000 & 1,000 & 1,000 & 1,000 & 1,000 & 1,000\\
    Spatial domain LR & $3e^{-4}$ & $3e^{-4}$ & $3e^{-4}$ & $3e^{-4}$ & $3e^{-4}$ & $3e^{-4}$ & $3e^{-4}$ & $3e^{-4}$ & $3e^{-4}$ & $3e^{-4}$ \\
    Spatial domain Batch size & 50 & 50 & 50 & 50 & 50 & 50 & 50 & 50 & 50 & 50 \\
    Spatial domain norm $C_t$ & 28 & 28 & 55 & 55 & 55 & 28 & 28 & 55 & 55 & 55 \\
    Spatial sample rate $q_t$ & 0.11 & 0.11 & 0.13 & 0.08 & 0.04 & 0.11 & 0.11 & 0.13 & 0.08 & 0.04 \\
    Spatial image amount $N_t$ & 50 & 50 & 50 & 500 & 500 & 50 & 50 & 50 & 500 & 500\\
    Frequency domain Epoch & 10 & 10 & 10 & 10 & 10 & 10 & 10 & 10 & 10 & 10\\
    Frequency domain LR & $3e^{-4}$ & $3e^{-4}$ & $3e^{-4}$ & $3e^{-4}$ & $3e^{-4}$ & $3e^{-4}$ & $3e^{-4}$ & $3e^{-4}$ & $3e^{-4}$ & $3e^{-4}$ \\
    Auxiliary Generator Epoch $M$ & 5 & 5  &  5 &  5 & 5 & 5 &5 & 5 &  5 &  5 \\
    Auxiliary Generator LR & 0.01 & 0.01  &  0.01 &  0.005 & 0.005 & 0.01 & 0.01  & 0.005 &  0.01 & 0.005  \\
    Auxiliary Generator Batch size & 100 & 100  &  100 & 500  & 500 & 100 & 100  & 500 &  100 & 500  \\
    Frequency domain Batch size $B_f$ & 256 & 256 & 256 & 256 & 256 & 256 & 256 & 256 & 256 & 256 \\
    Frequency feature dimension $K$& 10,000 & 10,000 & 10,000 & 10,000 & 10,000 & 10,000 & 10,000 & 10,000 & 10,000 & 10,000 \\
    Size of $D_s^f$ Image Dataset $N_f$ & 60,000 & 60,000 & 60,000 & 60,000 & 60,000 & 60,000 & 60,000 & 60,000 & 60,000 & 60,000\\
    Fine-tuning Epoch & 150 & 150 & 150 & 150 & 50 & 150 & 150 & 150 & 150 & 50\\
    Fine-tuning LR & $3e^{-4}$ & $3e^{-4}$ & $3e^{-4}$ & $3e^{-4}$ & $3e^{-4}$ & $3e^{-4}$ & $3e^{-4}$ & $3e^{-4}$ & $3e^{-4}$ & $3e^{-4}$ \\
    Fine-tuning Batch size & 4096 & 4096 & 4096 & 4096 & 4096 & 4096 & 4096 & 4096 & 4096 & 4096 \\
    Fine-tuning grad. norm & $1e^{-3}$ &  $1e^{-3}$ & $1e^{-3}$ & $1e^{-3}$ & $1e^{-3}$ & $1e^{-3}$ & $1e^{-3}$ & $1e^{-3}$ & $1e^{-3}$ & $1e^{-3}$\\
    Fine-tuning sample rate & $7.4e^{-2}$ &  $7.4e^{-2}$ & $9.1e^{-2}$ & $2.8e^{-2}$ & $1.5e^{-2}$ & $7.4e^{-2}$ &  $7.4e^{-2}$ & $9.1e^{-2}$ & $2.8e^{-2}$ & $1.5e^{-2}$ \\
    \bottomrule
\end{tabular}
}
\end{table*}
Given a batch size $B$, dataset size $N$, clipping hyper-parameter $C$, and noise variance $\sigma^2$, as defined in~\Cref{sub:dp}, the sampling ratio is $q = B / N$. The RDP privacy cost for a single SGM $\mathcal{Q}_{f,q,\sigma }\left( D \right)$ is given by~\Cref{eq:rdp_gamma}~\cite{sgm}.

\begin{theorem}[]
\label{eq:rdp_gamma}
Let $p_0 = \mathcal{N}(0, C^2 \sigma^2)$ and $p_1 = \mathcal{N}(1, C^2 \sigma^2)$ denote the probability density functions of two Gaussian distributions. A single SGM $\mathcal{Q}_{f,q,\sigma }\left( D \right)$ satisfies $(\alpha, \gamma_i)$-RDP for any $\gamma_i$ such that:
\begin{equation}
\gamma_i \geq D_\alpha\left( (1 - q) p_0 + q p_1 \, \| \, p_0 \right).
\end{equation}
\end{theorem}

This theorem enables the calculation of the per-step privacy bound $\gamma_i$ using the R\'{e}nyi divergence. To express this in terms of $(\varepsilon, \delta)$-DP, the following conversion applies.

\begin{theorem}[\textit{From $(\alpha, \gamma)$-RDP to $(\varepsilon, \delta)$-DP}~\cite{rdp}]
\label{eq:rdp_to_dp}
A mechanism $\mathcal{A}$ satisfying $(\alpha, \gamma)$-RDP also satisfies $(\varepsilon, \delta)$-DP for any $0 < \delta < 1$, where, $\varepsilon = \gamma + \frac{\ln (1/\delta)}{\alpha - 1}$.
\end{theorem}

Thus, by adjusting the noise variance $\sigma^2$, the final privacy cost $\varepsilon = \gamma + \frac{\ln (1/\delta)}{\alpha - 1}$ can be tailored to meet a target privacy budget $\varepsilon$.

\noindent \textbf{Privacy Composition.} 
When applying multiple differentially private mechanisms sequentially, their privacy costs must be combined to determine the overall privacy guarantee. For RDP, the composition theorem states that if $k$ mechanisms $\mathcal{Q}_1, \mathcal{Q}_2, \ldots, \mathcal{Q}_k$ each satisfy $(\alpha, \gamma_i)$-RDP, the composite mechanism satisfies $(\alpha, \gamma)$-RDP, where $\gamma = \sum_{i=1}^k \gamma_i.$

This additive property simplifies privacy accounting across multiple steps, such as in DP-SGD iterations or multi-stage frameworks in \toolname{}. The cumulative RDP cost $(\alpha, \gamma)$ can then be converted to $(\varepsilon, \delta)$-DP using~\Cref{eq:rdp_to_dp}.

\section{Proof of Sensitivity}
\label{apsec:proof}

\textbf{Proof of~\Cref{the:frequency}.} \textit{The query of frequency features of $D_s$ has global sensitivity $\Delta_{f}=1/N^\ast$. For any $\alpha>1$, incorporating noise $\mathcal{N}\left(0,{\sigma_f^2 \Delta_f^2} \mathbb{I} \right)$ into the mean random Fourier feature $\mu^p$ makes the query results satisfy $\left(\alpha, \gamma \right)$-RDP for some $\gamma$}.

\noindent \textit{Proof.} Let us assume two neighboring image datasets $D_s=\{h_i\}_{i=1}^N$ and $D_s'=\{h_i'\}_{i=1}^{N-1}$, where $h_i=h_i'$, $i\in [1,N-1]$. The sensitivity of the frequency features query is,
\begin{align*}
   \Delta_f = &\max \left\lVert \frac{1}{N^\ast} \sum_{i=1}^{N-1} \phi(h_i) 
  - \frac{1}{N^\ast} \sum_{i=1}^N \phi(h_i') 
 \right\lVert \\
 =&\max\left\lVert \frac{1}{N^\ast} \phi(h_N) \right\lVert \leq \frac{1}{N^\ast}. 
\end{align*}
To derive the third line from the second, we use the triangle inequality and the fact that $\|\phi(\cdot)\|=1$. We refer to implementation in DPImageBench~\cite{gong2025dpimagebench}, using $N^\ast \approx N$. We ignore the privacy budget caused by the dataset size estimation.

\vspace{1mm}
\noindent \textbf{Proof of~\Cref{the:meanImage}.} \textit{The query of spatial feature $h^\text{spat}$ has global sensitivity $\Delta_\text{spat}=C_t/B_t^\ast$. For any $\alpha>1$, incorporating noise $\mathcal{N}\left(0,{\sigma_t^2 \Delta_\text{spat}^2} \mathbb{I} \right)$ into the query result $h^\text{spat}$ makes the query results satisfy $\left(\alpha, \gamma \right)$-RDP for some $\gamma$.}

\noindent \textit{Proof.} We provide the proof of global sensitivity as follows. For any two neighboring image subsets $D_{s}^{\text{sub}}=\{h_i\}_{i=1}^{B_t}$ and $D_{s}^{\text{sub}'}=\{h_i\}_{i=1}^{B_t-1}$, the corresponding central images are $h^\text{spat}$ and 
$h^{\text{spat}'}$, we have,
\begin{equation}
    \begin{split}
        \Delta_\text{spat} =& {\left\| h^\text{spat} - h^{\text{spat}'} \right\|_2}\\
        =& {\left\| \frac{1}{B_t^*} \sum\nolimits_{i=1}^{B_t} {h_i} - \frac{1}{B_t^*}  \sum\nolimits_{i=1}^{B_t-1} {h_i} \right\|_2}\\
        =& {\left\| \frac{1}{B_t^*} h_{B_t} \right\|_2} \leq \frac{C_t}{B_t^*}. \\
    \end{split}
    \nonumber
\end{equation}
Therefore, we have $\Delta_\text{spat} \leq \frac{C_t}{B_t^*}$.

\vspace{1mm}
\noindent \textbf{Proof of~\Cref{the:frequency-disjoint}.} \textit{ Consider partitioning the dataset as $D_s = [D_s^1, D_s^2, \cdots, D_s^C]$, where the subsets are disjoint. By applying~\Cref{eq:freq_noise} to each subset, we generate noisy frequency features. The combined features have a dimension of $K\cdot C$, where $K$ is the predefined feature dimension and $C$ is the number of subsets. Because the subsets are disjoint, this mechanism has the same $\left(\alpha, \gamma \right)$-RDP guarantee as that in~\Cref{the:frequency}.}

\vspace{1mm}
\noindent \textit{Proof.} we categorize $D_s$ into $C$ disjoint subsets, $D_s = [D_s^1, D_s^2, \cdots, D_s^C]$. Let us assume two neighboring image datasets $D_s=\{h_i\}_{i=1}^N$ and $D_s'=\{h_i'\}_{i=1}^{N-1}$, where $h_i=h_i'$, $i\in [1,N-1]$. Besides, we assume the different image $h_i \in D_s^C$. The sensitivity of the frequency features query is,
\begin{align*}
   \Delta_f = &\max \left\lVert \frac{1}{N^\ast} \left[\sum_{i=1}^{|D_s^1|} \phi(h_i),\cdots,\sum_{i=1}^{|D_s^C|} \phi(h_i) \right] \right. \\
   & - \left. \frac{1}{N^\ast} \left[\sum_{i=1}^{|D_s^1|} \phi(h_i),\cdots,\sum_{i=1}^{|D_s^C|-1} \phi(h_i) \right] 
 \right\lVert \\
 =&\max\left\lVert \frac{1}{N^\ast} \left[ 0,\cdots, \phi(h_N) \right] \right\lVert \leq \frac{1}{N^\ast}. 
\end{align*}
The derivation from the second line to the third leverages the triangle inequality and the property that $\|\phi(\cdot)\|=1$. Similar to the implementations in~\Cref{the:frequency}, we utilize the approximation $N^\ast \approx N$ and disregard the privacy budget incurred by this dataset size estimation.

\section{Implementation Details}
\label{apsec:id}

This section introduces the baselines (including existing methods and variants of \toolname), details of selected metrics, and hyper-parameter settings of \toolname.

\subsection{Beselines}
\label{apsubsec:baselines}

In our experiments, we implement all baselines using the open-source DP image synthesis benchmark DPImageBench~\cite{gong2025dpimagebench}. We introduce baselines as follows:

\begin{itemize}[leftmargin=*]

\item \textbf{DP-MERF~\cite{dp-merf}:} DP-MERF leverages random feature representations of kernel mean embeddings with the Maximum Mean Discrepancy (MMD)~\cite{MMD} to minimize the distributional distance between real and synthetic data.

\item \textbf{DP-NTK~\cite{dp-ntk}:} DP-NTK uses Neural Tangent Kernels~\cite{ntks} to represent images, using the gradient of the neural network function as a feature map to extract perceptual features from original images for DP synthesis.

\item \textbf{DP-Kernel~\cite{dp-kernel}:} DP-Kernel uses functional RDP to privatize the loss function of the data generator within a reproducing kernel Hilbert space, enabling DP image synthesis.

\item \textbf{GS-WGAN~\cite{gs-wgam}:} GS-WGAN perturbs only the discriminator’s feedback to guide the generator, ensuring DP. By applying the chain rule, it decomposes the generator’s gradient into an \textit{upstream gradient} (the discriminator’s output with respect to the generated image) and a \textit{local gradient} (the generated image with respect to the generator’s parameters), perturbing only the upstream gradient for DP, as the discriminator solely accesses sensitive images. 

\item \textbf{DP-GAN~\cite{dpgan}:} DP-GAN trains the discriminator network on sensitive images using DP-SGD~\cite{dpsgd}, ensuring that the discriminator weights satisfy differential privacy (DP). The generator's weight updates depend solely on the discriminator. By the post-processing property of DP~\cite{dpbook}, the generator also satisfies DP.

\item \textbf{DPDM~\cite{dpdm}:} DPDM trains diffusion models on sensitive images using DP-SGD~\cite{dpsgd}. It introduces noise multiplicity, a modification to DP-SGD, to mitigate the adverse effects of injected noise on gradients.

\item \textbf{DP-FETA~\cite{dp-feta}:} DP-FETA uses a two-stage training process: (1) It extracts basic features (e.g., outlines, colors) from sensitive images using `central images' -- derived from central tendency measures (e.g., mean, mode). (2) Fine-tuning on sensitive images with DP-SGD~\cite{dpsgd}.

\end{itemize}

Then, we introduce the variants of \toolname{} studied in~\Cref{sec:rq2} as follows.

\begin{itemize}[leftmargin=*]
\item \textbf{\toolname-No-Auxiliary}. This method means using the synthesizer, diffusion model $e_\theta$, to learn the frequency feature directly. Similar to~\Cref{eq:mmd}, at each training iteration, this method uses the diffusion model to generate a batch of synthetic images, and then it optimizes the model parameters to minimize the difference between the frequency features of synthetic images and sensitive images. The formal objective is
\begin{equation}
\label{eq:mmd_no_aux}
    \mathcal{L}(\theta) = \|\tilde{\mu} - \mu^s(\text{Sampler}(e_\theta, z))\|_2,
\end{equation}
where $z$ is a batch of random noise. `Sampler' uses $e_\theta$ to denoise $z$ into a batch of synthetic images through multiple steps of denoising. Although we can obtain the denoised images before the final step, these images are still too noisy and can hardly contain useful frequency features. Therefore, we use the clean image at the final step to calculate its frequency feature. This paper uses the DDIM Sampler~\cite{ddim} with a sampling step of 50.

\item \textbf{\toolname-DM-Auxiliary}. This method means using diffusion models as the auxiliary generator, and other components keep consistent with those in \toolname.
\end{itemize}

\subsection{Dataset Details}
\label{apsec:dataset}

In this paper, we conduct experiments on five image datasets that are widely used in prior works~\cite{li2023PrivImage,gong2025dpimagebench,dp-feta,pe3,dpldm,dplora} to verify the effectiveness of \toolname compared with baselines: {\tt MNIST}, {\tt FashionMNIST}~\cite{fmnist} ({\tt F-MNIST}), {\tt CIFAR-10}~\cite{cifar10}, {\tt CelebA}~\cite{celeba}, and {\tt Camelyon}~\cite{camelyon1}.

The \texttt{MNIST} dataset consists of 70,000 grayscale images of handwritten digits (0–9). \texttt{F-MNIST} includes 70,000 images of 10 distinct fashion products. \texttt{CIFAR-10} comprises 10 classes of natural images, consisting of 60,000 images.
Compared to \texttt{MNIST}, \texttt{F-MNIST}, and \texttt{CIFAR-10}, \texttt{CelebA} and \texttt{Camelyon} are more sensitive datasets. \texttt{CelebA} contains over 202,599 facial images of 10,177 celebrities, each annotated with 40 attributes; following prior work~\cite{dpsda,dp-feta,gong2025dpimagebench}, we use the `Gender' attribute to classify images as male or female. \texttt{Camelyon} includes 455,954 histopathological image patches of human tissue, labeled based on the presence of at least one tumor cell pixel. As shown in~\Cref{tab:datainfo}, all datasets are split into training, validation, and test sets. For \texttt{CelebA} and \texttt{Camelyon}, all images are center-cropped and resized to $32 \times 32$ pixels.

\begin{table*}[!t]
\renewcommand{\arraystretch}{1.1}
\setlength{\tabcolsep}{5.5pt}
\small
    \centering
    \caption{The IS, Precision, Recall, and FLD of \toolname{} and seven baselines on {\tt MNIST}, {\tt F-MNIST}, {\tt CIFAR-10}, {\tt CelebA} and {\tt Camelyon} with $\varepsilon=\{1,10\}$. `\toolnamef' learns frequency domain knowledge before spatial domain knowledge. `Pre.' and `Re.' are the abbreviations of `Precision' and `Recall.' Note that IS only makes sense for natural images, such as {\tt CIFAR-10}; we include it for all datasets for completeness.}
    \label{tab:add_metrics}
    \resizebox{1.0\textwidth}{!}{
    \renewcommand{\arraystretch}{1.1}
    \begin{tabular}{l|rc|rc|rc|rc|rc|rc|rc|rc|rc|rc}
    \toprule
    \multirow{3}{*}{\textbf{Method}} & \multicolumn{10}{c|}{$\varepsilon=1$} & \multicolumn{10}{c}{$\varepsilon=10$}\\
    \Xcline{2-21}{0.5pt}
    & \multicolumn{2}{c|}{{\tt MNIST}} & \multicolumn{2}{c|}{{\tt F-MNIST}} & \multicolumn{2}{c|}{{\tt CIFAR-10}} &  \multicolumn{2}{c|}{{\tt CelebA}} & \multicolumn{2}{c|}{{\tt Camelyon}} & \multicolumn{2}{c|}{{\tt MNIST}} & \multicolumn{2}{c|}{{\tt F-MNIST}} & \multicolumn{2}{c|}{{\tt CIFAR-10}} & \multicolumn{2}{c|}{{\tt CelebA}} & \multicolumn{2}{c}{{\tt Camelyon}} \\
    \Xcline{2-21}{0.5pt}
     & \centering IS & Pre. & IS & Pre. & IS & Pre. & IS & Pre. & IS & Pre. & IS & Pre. & IS & Pre. & IS & Pre. & IS & Pre. & IS & Pre. \\
    \hline
    DP-MERF & 2.55 & 0.02 & 2.93 & 0.08 & 2.98 & 0.29 & 3.90 & 0.44 & 1.43 & 0.09 & 2.64 & 0.03 &  2.86 & 0.08 & 3.06 & 0.25 & 3.61 & 0.08 & 2.45 & 0.12 \\
    DP-NTK & 1.60 & 0.26 &  2.90 & 0.07  & 1.23 & 0.00 & 1.57 & 0.00 & 1.52 & 0.00 & 2.18 & 0.09 & 3.05 & 0.04 & 1.59 & 0.01 & 3.21 & 0.05 & 1.67 & 0.00 \\
    DP-Kernel & 2.15 & 0.28 & 3.53 & 0.20 & 2.99 & 0.05 & 3.16 & 0.07 & 2.30 & 0.00 & 2.19 & 0.24 & 3.45 & 0.23 & 3.69 & 0.18 & 2.90 & 0.11 & 3.21 & 0.01 \\
    GS-WGAN & 2.19 & 0.05 & 2.93 & 0.13 & 1.68 & 0.18 & 1.00 & 0.00 & 1.23 & 0.00 & 2.37 & 0.13 & 2.95 & 0.17 & 2.34 & 0.17 & 1.66 & 0.03 & 1.45 & 0.04 \\
    DP-GAN & 1.52 & 0.16 & 3.51 & 0.24 & 1.83 & 0.26 & 3.31 & 0.30 & 1.76 & 0.25 & 2.06 & 0.19 & 3.60 & 0.17 & 2.65 & 0.67 & 2.28 & 0.62 & 1.84 & 0.37 \\
    DPDM & 2.23 & 0.61 & 3.43 & 0.27 & 1.82 & 0.63 & 2.63 & 0.42 & 1.53 & 0.66 & 2.07 & 0.63 & 3.92 & 0.54 & 3.12 & 0.59 & 2.23 & 0.60 & 1.65 & 0.74 \\
    DP-FETA & 2.02 & 0.47 & 3.75 & 0.39 & 2.46 & 0.57 & 2.38 & 0.41 & 1.62 & 0.68 &  2.09 & 0.66  & 3.93 & 0.60 & 3.30 & 0.63 & 2.30 & 0.63 & 1.69 & 0.72 \\
    \hline
    \rowcolor{gray0} \toolnamef & 2.03 & 0.52 & 3.77 & 0.39 & 2.48 & 0.62 & 2.32 & 0.38 & 1.59 & 0.68 & 2.07 & 0.68 & 3.94 & 0.62 & 3.92 & 0.58 & 2.36 & 0.61 & 1.71 & 0.72 \\
     \rowcolor{gray0} \toolname & 2.04 & 0.54 & 3.76 & 0.41 & 2.81 & 0.65 & 2.49 & 0.51 & 1.66 & 0.71 &  2.08  & 0.69  & 3.97 & 0.61 & 4.46 & 0.48 & 2.36 & 0.61 & 1.72 & 0.73 \\
    \bottomrule
    \toprule
    \multirow{3}{*}{\textbf{Method}} & \multicolumn{10}{c|}{$\varepsilon=1$} & \multicolumn{10}{c}{$\varepsilon=10$}\\
    \Xcline{2-21}{0.5pt}
    & \multicolumn{2}{c|}{{\tt MNIST}} & \multicolumn{2}{c|}{{\tt F-MNIST}} & \multicolumn{2}{c|}{{\tt CIFAR-10}} &  \multicolumn{2}{c|}{{\tt CelebA}} & \multicolumn{2}{c|}{{\tt Camelyon}} & \multicolumn{2}{c|}{{\tt MNIST}} & \multicolumn{2}{c|}{{\tt F-MNIST}} & \multicolumn{2}{c|}{{\tt CIFAR-10}} & \multicolumn{2}{c|}{{\tt CelebA}} & \multicolumn{2}{c}{{\tt Camelyon}} \\
    \Xcline{2-21}{0.5pt}
     & \centering Re. & FLD & Re. & FLD & Re. & FLD & Re. & FLD & Re. & FLD & Re. & FLD & Re. & FLD & Re. & FLD & Re. & FLD & Re. & FLD \\
    \hline
    DP-MERF & 0.03 & 37.9 & 0.00 & 27.6 & 0.00 & 38.9 & 0.00 & 28.8 & 0.00 & 56.3 & 0.03 & 34.9 & 0.01 & 29.2 & 0.00 & 32.1 & 0.00 & 17.1 & 0.00 & 50.5 \\
    DP-NTK & 0.00 & 68.1 & 0.00 & 40.3 & 0.00 & 51.9 & 0.00 & 62.2 & 0.00 & 60.1 & 0.08 & 25.5 & 0.00 & 36.4 & 0.00 & 41.7 & 0.00 & 30.6 & 0.00 & 52.7 \\
    DP-Kernel & 0.04 & 17.1 & 0.01 & 20.1 & 0.00 & 30.6 & 0.00 & 15.8 & 0.00 & 45.1 & 0.02 & 17.8 & 0.01 & 21.3 & 0.00 & 27.2 & 0.00 & 14.1 & 0.01 & 38.0 \\
    GS-WGAN & 0.00 & 27.0 & 0.00 & 28.5 & 0.00 & 35.8 & 0.00 & 77.0 & 0.00 & 80.8 & 0.01 & 25.4 & 0.00 & 28.1 & 0.00 & 31.1 & 0.00 & 43.2 & 0.01 & 64.2 \\
    DP-GAN & 0.00 & 31.9 & 0.02 & 24.4 & 0.00 & 27.6 & 0.00 & 14.8 & 0.12 & 27.3 &  0.22 & 15.0  & 0.02 & 23.9 & 0.01 & 22.5 & 0.05 & 3.9 & 0.14 & 3.01 \\
    DPDM & 0.15 & 10.4 & 0.12 & 20.4 & 0.00 & 28.6 & 0.00 & 23.8 & 0.18 & 26.9 &  0.73 & 3.3 & 0.38 & 6.6 & 0.04 & 19.4 & 0.15 & 4.5 & 0.29 & -2.7 \\
    DP-FETA & 0.62 & 8.1 & 0.26 & 11.4 & 0.01 & 23.2 & 0.00 & 11.2 & 0.27 & -3.1 & 0.77 & 2.7  & 0.43 & 4.6 & 0.02 & 16.8 & 0.17 & 3.5 & 0.31 & -4.4 \\
    \hline
    \rowcolor{gray0} \toolnamef & 0.67 & 6.9 & 0.28 & 11.3 & 0.01 & 22.3 & 0.02 & 14.1 & 0.23 & -2.9 & 0.76 & 2.3 & 0.42 & 4.5 & 0.10 & 17.0 & 0.22 & 3.3 & 0.32 & -4.7 \\
     \rowcolor{gray0} \toolname & 0.69 & 5.7 & 0.29 & 10.7 & 0.02 & 20.4 & 0.10 & 9.0 & 0.30 & -3.5 & 0.77 & 2.1 & 0.46 & 4.4 & 0.17 & 16.2 & 0.23 & 3.2 & 0.34 & -5.0 \\
    \bottomrule
\end{tabular}
}
\end{table*}

\subsection{Details of Metrics}
\label{apsubsec:MetricDetails}

We use `entropy' and `texture complexity' to evaluate the complexity of images. We elaborate on them as follows. 
\begin{itemize}[leftmargin=*]
\item \textbf{Entropy:} Entropy measures the level of uncertainty in a set of data. A high entropy value means the pixel values are highly random and diverse, indicating a greater amount of information and, therefore, more complexity~\cite{sonka2013image}. Calculating image entropy involves these steps: (1) \textit{Grayscale Conversion}: If the image is in color, we first convert it to grayscale so that each pixel has a single value (0-255). (2) \textit{Histogram:} We create a histogram of the image's pixel values, which gives a probability distribution $p_k$. Here, $p_k$ is the number of pixels with a value of $k$. \textit{Shannon Entropy:} We then use the Shannon Entropy~\cite{sonka2013image} to calculate the entropy value: $H = - \sum_{k=0}^{K-1} p_j \log (p_k).$
\item \textbf{Texture complexity:} Texture complexity evaluates an image based on the richness of its visual details, patterns, and structures. It captures the kind of complexity the human visual system perceives, such as the number of edges and the regularity of patterns. An image with more details and irregular patterns has a higher texture complexity~\cite{gonzalez2009digital}.
\end{itemize}

We evaluate the fidelity and utility of the synthetic dataset using two metrics: Fr´echet Inception Distance (FID) and downstream classification accuracy (Acc). We generate 60,000 synthetic images for evaluations.
\begin{itemize}[leftmargin=*]
\item  \textbf{FID:} The FID is widely used to assess the quality of synthetic images generated by models such as those in~\cite{ddpm,biggan,dpdm}. A lower FID indicates higher-quality images that closely resemble the real dataset. To compute the FID, we utilize the pre-trained Inception V3 model~\cite{inceptionv3} from GitHub to extract features or classify images.\footnote{\url{https://github.com/mseitzer/pytorch-fid/releases/download/fid_weights/pt_inception-2015-12-05-6726825d.pth}}

\item \textbf{Acc:} We evaluate the utility of synthetic images for image classification tasks. Following DPImageBench~\cite{gong2025dpimagebench}, we train three classification models—ResNet~\cite{resnet}, WideResNet~\cite{zagoruyko2017wideresidualnetworks}, and ResNeXt~\cite{resnext}—using synthetic images and assess their accuracy on a sensitive test dataset. The highest accuracy among these classifiers is reported. Referring to the implementations in DPImageBench~\cite{gong2025dpimagebench}, as the validation set is sensitive data, we use the `Report Noisy Max' algorithm~\cite{report-noisy-max} to select the optimal classifier checkpoint across all epochs for classifiers. Then, we report the accuracy of this classifier on the test set. This procedure prevents inflated accuracy due to train-test overlap or DP violations during hyper-parameter tuning.
\end{itemize}

\subsection{Hyper-parameter Settings of \toolname}
\label{apsubsec:hyper}

This section introduces the hyper-parameter settings of \toolname. \Cref{tab:hyper-parameter} summarizes the hyper-parameter settings used in our experiments for \toolname{} across different datasets and privacy levels ($\epsilon=1$ and $\epsilon=10$). As described in~\Cref{sec:methodology}, \toolname{} follows a staged training approach consisting of three phases: spatial domain learning, frequency domain learning, and DP-SGD fine-tuning. \Cref{tab:hyper-parameter} reports the noise scales ($\sigma_f$, $\sigma_t$) used for DP, the sample rate ($q_c$), and the number of central images updates ($N_c$). It also lists the epoch counts, learning rates (LR), and batch sizes used in each training phase.
We observe that while most hyper-parameters remain consistent across datasets, such as batch size and LR, we adapt noise scales and some sampling strategies depending on data complexity and privacy requirements. For example, higher noise is applied under stricter privacy constraints ($\epsilon=1$), especially on more complex datasets like \texttt{Camelyon}.

\section{Additional Analysis}
\label{apsec:add_analysis}

We introduce the fidelity analysis, the fractions of privacy budget spent in \toolname, and the synthetic performance when using the PRV~\cite{PRV} privacy budget accounting method.

\subsection{Fidelity Analysis}
\label{apsubsec:fidelity}

\begin{table}[!t]
\small
    \centering
    \caption{Noise scale of DP-SGD in \toolname{} under different privacy allocation strategies. }
    \label{tab:dpsgd}
    \setlength{\tabcolsep}{3.2mm}{
    \resizebox{0.50\textwidth}{!}{
    \renewcommand{\arraystretch}{1.2}
    \begin{tabular}{l|ccccc}
    \toprule
    \textbf{Noise scale}  & $\sigma_t = 2$  & $\sigma_t = 5$ & $\sigma_t = 10$  & $\sigma_t = 20$  & $\sigma_t = 30$  \\
    \hline
    $\sigma_f = 20$ & 152.4&50.98&39.56&37.67&36.22  \\
    $\sigma_f = 26$ & 16.74&16.00&15.51&15.39&15.29 \\
    $\sigma_f = 42$ & 16.33&15.65&15.19&15.07&14.98 \\
    $\sigma_f = 61$ & 16.25&15.57&15.12&15.01&14.91  \\
    $\sigma_f = 115$ & 16.23&15.55&15.10&14.99&14.90 \\
    \bottomrule
\end{tabular}
}}
\end{table}

Referring to fidelity metrics implemented in DPImageBench~\cite{gong2025dpimagebench}, we leverage more fidelity metrics to evaluate the synthetic images. Precision and recall~\cite{precision&recall} provide separate scores to offer a clearer understanding of whether a model excels at generating high-quality images (precision) or diverse images (recall). Moreover, \toolname{} also incorporates Feature Likelihood Divergence (FLD)~\cite{fld}, which evaluates the novelty of the synthetic dataset. We elaborate on these metrics as follows.

\vspace{1mm}
\noindent \textbf{Inception Score (IS):} The IS leverages the Inception v3 network~\cite{inceptionv3} to classify generated images into predefined categories. A higher IS reflects synthetic images that are both realistic (classified with high confidence) and diverse (spanning multiple classes).

\begin{table*}[!t]
\small
    \centering
    \caption{RDP cost ratios (\%) of spatial features / frequency features / DP-SGD in \toolname{} using various privacy allocation strategies, under the $\epsilon=1.0$.}
    \label{tab:rdpRatio}
    \setlength{\tabcolsep}{4.5mm}{
    \resizebox{1.0\textwidth}{!}{
    \renewcommand{\arraystretch}{1.2}
    \begin{tabular}{l|ccccc}
    \toprule
    \textbf{Noise scale } & $\sigma_t = 2$  & $\sigma_t = 5$ & $\sigma_t = 10$  & $\sigma_t = 20$  & $\sigma_t = 30$  \\
    \hline
    $\sigma_f = 20$ & 80.83 / 6.15 / 13.02 & 5.24 / 4.98 / 89.78 & 1.22 / 4.99 / 93.8 & 0.3 / 4.94 / 94.77 & 0.13 / 4.94 / 94.92  \\
    $\sigma_f = 26$ & 80.07 / 3.6 / 16.33 & 5.25 / 2.95 / 91.8 & 1.22 / 2.95 / 95.83 & 0.3 / 2.92 / 96.78 & 0.13 / 2.93 / 96.94 \\
    $\sigma_f = 42$ & 80.04 / 1.38 / 18.58 & 5.24 / 1.13 / 93.63 & 1.21 / 1.13 / 97.66 & 0.29 / 1.12 / 98.59 & 0.13 / 1.12 / 98.75 \\
    $\sigma_f = 61$ & 80.63 / 0.66 / 18.71 & 5.17 / 0.53 / 94.3 & 1.2 / 0.53 / 98.28 & 0.13 / 0.53 / 99.34 & 0.13 / 0.53 / 99.34  \\
    $\sigma_f = 115$ & 81.01 / 0.19 / 18.8 & 5.19 / 0.15 / 94.66 & 1.2 / 0.15 / 98.65 & 0.3 / 0.15 / 99.55 & 0.13 / 0.15 / 99.72 \\
    \bottomrule
\end{tabular}
}}
\end{table*}

\vspace{1mm}
\noindent \textbf{Precision and Recall:} We extract features from sensitive and synthetic images using the Inception v3 network. For precision, we identify the nearest sensitive image in the feature space for each synthetic image. A synthetic image is deemed a `true positive' if its distance to the closest sensitive image is below a predefined threshold. Precision is computed as $\frac{\text{True Positives}}{\text{Total Synthetic Images}}$. For recall, we find the $k$-neighboring-nearest (default $k=4$) synthetic images for each sensitive image, labeling a pair as a `true positive' if their distance is below the threshold. Recall is calculated as $\frac{\text{True Positives}}{\text{Total Sensitive Images}}$. Higher precision and recall are desirable~\cite{precision&recall}.

\begin{table}[!t]
\renewcommand{\arraystretch}{1.1}
\setlength{\tabcolsep}{5.5pt}
\small
    \centering
    \caption{FID and Acc (\%) of \toolname{} on five sensitive datasets with $\varepsilon=10$ using RDP and PRV privacy budget accounting methods.}
    \label{tab:prv_rdp}
    \resizebox{0.49\textwidth}{!}{
    \renewcommand{\arraystretch}{1.2}
    \begin{tabular}{l|cc|cc|cc|cc|cc}
    \toprule
    \multirow{2}{*}{\textbf{Method}} & \multicolumn{2}{c|}{{\tt MNIST}} & \multicolumn{2}{c|}{{\tt F-MNIST}}  & \multicolumn{2}{c|}{{\tt CIFAR-10}}  &  \multicolumn{2}{c|}{{\tt CelebA}} & \multicolumn{2}{c}{{\tt Camelyon}} \\
    \Xcline{2-11}{0.5pt}
     & \centering FID & Acc & FID & Acc & FID & Acc & FID & Acc & FID & Acc \\
    \hline
    RDP  &  2.5 & 98.6 & 11.6  & 88.1 & 69.0 & 47.0 & 20.0 & 95.2 & 20.3 & 84.3 \\
    PRV & 2.6 & 98.5 & 10.9 & 88.1 & 70.3 & 46.9 & 18.6 & 95.1 & 21.9 & 86.0 \\
    \bottomrule
\end{tabular}
}
\end{table}

\vspace{1mm}
\noindent \textbf{Fréchet Leakage Distance (FLD):} Building on FID, FLD further evaluates the novelty of synthetic images. Lower values indicate better performance, with negative values suggesting the synthetic dataset surpasses the quality of the original synthetic images to some extent.

\Cref{tab:add_metrics} presents a comprehensive evaluation of \toolname{}, \toolnamef{}, and seven baseline methods across five investigated sensitive datasets under privacy budgets $\varepsilon=1$ and $\varepsilon=10$, using IS, Precision, Recall, and FLD. Under all cases, \toolname{} attains the best FLD scores compared to baselines. It is important to note that Precision and Recall offer complementary insights independent of the FID score. High Precision indicates generated images are high-quality and resemble real images, whereas high Recall reflects the model’s ability to capture the full diversity of the original dataset. However, a high score in one does not guarantee a low FID if the other is poor." For example, in the case of GS-WGAN with {\tt CIFAR-10} under $\epsilon=10$, the precision is 0.67, the best among all methods. However, its recall is as low as 0.00, and the FID is as high as 194.4, indicating poor overall synthetic image quality.

\toolname{} achieves the highest IS (e.g., 4.46 on {\tt CIFAR-10} at $\varepsilon=10$), Precision (0.71 on {\tt Camelyon} at $\varepsilon=1$), and Recall (0.77 on {\tt MNIST} at $\varepsilon=10$), alongside the lowest FLD (e.g., -5.0 on {\tt Camelyon} at $\varepsilon=10$), indicating superior realism, diversity, and novelty. \toolnamef{}, which prioritizes frequency domain learning, performs slightly worse (e.g., FLD 6.9 vs. 5.7 on {\tt MNIST} at $\varepsilon=1$). Notably, \toolname{} excels under strict privacy ($\varepsilon=1$), though low Recall on {\tt CIFAR-10} and {\tt CelebA} suggests room for improving diversity on complex datasets. These results, consistent with~\Cref{tab:rq1} and~\Cref{fig:fid_curve}, present \toolname{}’s robustness in DP image synthesis.

\subsection{Privacy Allocation Strategy}
\label{apsubsec:privacy_allocation}

\Cref{tab:dpsgd,tab:rdpRatio} present the noise scales and RDP budget ratios, respectively, of three processes in \toolname{} (spatial feature extraction, frequency feature extraction, and DP-SGD, as described in~\Cref{sec:methodology}) under different privacy allocation strategies. For each process, a larger noise scale indicates that more noise is injected, thereby consuming less privacy budget to achieve the same DP. Therefore, the privacy budget allocation ratio across the three processes increases as the corresponding noise scale decreases. We introduce the calculation of the RDP budget ratio of each process. Each process of \toolname{} can be viewed as a sub-sampled Gaussian Mechanism (SGM)~\cite{sgm}, as introduced in~\Cref{app:supp_dp}. Thus, the three processes of \toolname{} can be viewed as three SGMs, $\mathcal{Q}_t$, $\mathcal{Q}_f$, and $\mathcal{Q}_{d}$.

\Cref{app:supp_dp} explains that RDP has a nice linear composability property. For three mechanisms $\mathcal{Q}_t$, $\mathcal{Q}_f$, and $\mathcal{Q}_{d}$ satisfying ($\alpha, \gamma_t$)-RDP, ($\alpha, \gamma_f$)-RDP, and ($\alpha, \gamma_{d}$)-RDP, respectively, the composition ($\mathcal{Q}_t$, $\mathcal{Q}_f$, $\mathcal{Q}_{d}$) satisfies ($\alpha, \gamma_t+\gamma_f+\gamma_{d}$)-RDP. According to the spatial and frequency feature noise $\sigma_t$ and $\sigma_f$, we can determine the optimal $\alpha^*$ to obtain the minimal $\gamma_t^*+\gamma_f^*+\gamma_{d}^*$. It is noticed that the noise scale of DP-SGD is calculated by adaptively adjusting as introduced in Section~\ref{subsec:privacy_analy} to meet the given privacy budgets. The privacy budget ratio of each process (e.g., spatial features) can be calculated using the equation, $\frac{\gamma_t^*}{\gamma_t^*+\gamma_f^*+\gamma_{d}^*}$. These ratios offer critical insight into how privacy budgets are distributed across the pipeline.

\subsection{Privacy Accounting Using PRV}
\label{apsubsec:PRV}

In this paper, we use RDP to track the privacy consumption of our method for a \textit{fair comparison} with existing approaches, as RDP has become a widely adopted standard in privacy accounting for DP machine learning~\cite{dp-feta,gong2025dpimagebench,li2023PrivImage}. We also explore the use of another privacy accounting method, Privacy Random Variable (PRV)~\cite{PRV}. Due to the similar privacy analysis frameworks of PRV and RDP, both of which support moment accounting and composition over multiple training steps, we can seamlessly integrate the PRV accountant into \toolname{}. Specifically, we used the PRV accountant provided by the Opacus\footnote{\url{https://opacus.ai/}} library to replace the RDP accountant originally used in \toolname.

\Cref{tab:prv_rdp} presents the FID and accuracy of \toolname{} on five sensitive datasets with $\epsilon = 10$, comparing the RDP and PRV privacy budget accounting methods. From this table, we can see that the synthetic performance achieved by RDP and PRV is similar. The privacy accounting methods have a small impact on the final synthetic performance. Specifically, the average changes of FID and Acc between these two privacy accounting methods are 0.18 $(=|((2.5-2.6) + (11.6-10.9) + (69.0-70.3) + (20.0-18.6) + (20.3-21.9))|/5 )$ and 0.28\% $(=|((98.6-98.5) + (88.1-88.1) + (47.0-46.9) + (95.2-95.1) + (84.3-86.0))|/5 \times 100\%)$, respectively. Although PRV theoretically provides a tighter analysis of privacy loss by modeling the exact distribution of privacy losses across iterations, its practical benefit for \toolname{} is marginal.

\subsection{Performance on High-Resolution Images}
\label{apsubsec:high_resolution}

Table~\ref{tab:resulotion} presents the Acc and FID of synthetic images generated from the {\tt CelebA} dataset ($\epsilon=10$) across three resolutions: $32\times 32$, $64 \times 64$, and $128 \times 128$. We observe that \toolname{} achieves the best performance compared to baselines. In particular, at a $128 \times 128$ resolution, \toolname{} achieves an Accuracy of 73.4\% and an FID of 146.3, outperforming the state-of-the-art DP-FETA, which achieves 69.3\% and 173.6, respectively. Although \toolname{} achieves the best performance, improving the utility of synthesizers for high-resolution images, without relying on public resources, remains a promising direction for future research.

\begin{table}[!t]
\small
    \centering
    \caption{The Acc $(\%)$ and FID of synthetic images using {\tt CelebA} under $\epsilon=10$, with varying image resolutions. }
    \label{tab:resulotion}
    \setlength{\tabcolsep}{3.2mm}{
    \resizebox{0.48\textwidth}{!}{
    \begin{tabular}{l|cc|cc|cc}
    \toprule
    \multirow{2}{*}{Algorithm} & \multicolumn{2}{c|}{{\tt $32 \times 32$}} & \multicolumn{2}{c|}{{\tt $64 \times 64$}}  & \multicolumn{2}{c}{{\tt $128 \times 128$}} \\
    \cline{2-7}
    & Acc  & FID & Acc & FID  & Acc & FID \\
    \hline
    DP-MERF & 81.2 & 147.9 & 61.7  & 381.2 & 61.4 & 311.2 \\
    DP-NTK & 64.2 & 227.8 & 64.2 & 299.9 & 62.9 & 272.1 \\
    DP-Kernel & 83.7 & 128.8 & 73.5 & 272.4 & 53.7 & 359.2 \\
    GS-WGAN & 61.5 & 290.0 & 61.8 & 433.6 & 61.4 & 383.2 \\
    DP-GAN & 89.2 & 31.7 & 61.8 & 395.0 & 39.9 & 320.2 \\
    DPDM   & 91.8 & 28.8 & 78.7 &  106.7 & 71.1 & 210.8 \\
    DP-FETA   & 94.2 & 24.8 & 82.6 & 89.4 & 69.3 & 173.6 \\
    \hline
    \rowcolor{gray0} \toolname & \textbf{95.2} & \textbf{20.0} & \textbf{90.4} & \textbf{72.7} & \textbf{73.4} & \textbf{146.3} \\
    \bottomrule
\end{tabular}}}
\end{table}

\section{Limitations}
\label{subsec:limit}

\toolname{} consumes privacy budget across three stages: (1) warming up with spatial features, (2) warming up with frequency features, and (3) fine-tuning synthesizers on original datasets using DP-SGD. As discussed in~\Cref{sec:impact_hyper}, the synthetic performance of \toolname{} is highly sensitive to privacy budget allocations. Suboptimal privacy budget allocation plans may greatly damage the synthetic performance of \toolname, while exhaustively evaluating all possible allocation strategies is computationally prohibitive. Future work will explore highly efficient methods to identify optimal privacy budget allocation strategies. Current DP image synthesis methods overlook the privacy cost associated with hyper-parameter tuning~\cite{gong2025dpimagebench,dpsda,dpldm}.
Our method follows them and similarly does not account for this cost.

This paper introduces two types of training shortcuts; however, other shortcuts could further improve DP image synthesis performance. We plan to explore more efficient features as training shortcuts in the future work.

\balance
\section{Related Work}
\label{suppsec:related}
This section provides a complement to the related works discussed in~\Cref{sec:related}. We discuss previous DP image synthesis methods based on whether the methods leverage public resources or not. The public resources involve public datasets or pre-trained models~\cite{gong2025dpimagebench}. Public image datasets refer to the datasets released on open-source platforms (e.g., {\tt ImageNet}~\cite{imagenet}) without privacy concerns. Pre-trained models are accessible via cloud-based services (e.g., DALL-E~\cite{dalle2}) or local software libraries (e.g., Stable Diffusion~\cite{labelembedding}).

\vspace{1mm}
\noindent \textbf{Without using Public Resources.} These methods train the synthesizer solely using the sensitive image dataset, without relying on external public resources~\cite{dp-merf,dp-feta,pearl,dp-ntk}. A group of methods proposes adding noise to the high-level feature for DP, e.g., random Fourier features~\cite{dp-mepf,pearl}, empirical neural tangent kernels~\cite{dp-ntk}, of sensitive images, and training a synthesizer to ensure synthetic images approach the noisy feature. Alternatively, rather than perturbing the features, Yang et al.~\cite{dp-kernel} introduce DP-Kernel, which directly injects Gaussian noise into the loss function that quantifies the divergence between the distributions of sensitive and synthetic images under DP constraint.

Although the aforementioned methods have low computational requirements, they present suboptimal synthetic performance on complex datasets, such as {\tt CelebA}~\cite{celeba}. Abadi et al.~\cite{dpsgd} introduce DP-SGD, which adds noise to training gradients to achieve DP. Various works use DP-SGD to train deep generative models on sensitive datasets for addressing DP synthesis of complex images~\cite{dpgan,g-pate,dpdm,dpldm,dp-diffusion,li2023PrivImage,dplora}. Among these approaches, utilizing diffusion models as synthesizers achieves SOTA synthetic performance. For instance, Dockhorn et al.~\cite{dpdm} introduced DPDM, which reduces the variance of noisy gradients by reusing sensitive images during training to learn denoising across multiple steps.
Building on DPDM, Li et al.~\cite{dp-feta} introduce DP-FETA, which uses a two-stage training approach: (1) extracting fundamental features (e.g., outlines and colors) from sensitive images, and (2) capturing comprehensive feature representations of sensitive images. This progressive `from easy to hard' learning framework further improves synthetic performance.

This paper follows the `from easy to hard' framework proposed in DP-FETA~\cite{dp-feta} and investigates the refinement of training shortcuts to enhance the performance of DP image synthesis.

\vspace{1mm}
\noindent \textbf{Using Public Dataset.} Several methods suggest using a publicly available dataset to initially train a data synthesizer, followed by fine-tuning models with sensitive images using DP-SGD, to improve synthetic performance~\cite{dp-mepf,chen2022dpgen,dp-diffusion,dplora,dpldm,yin2022practical,lin2020using}. This paradigm represents the predominant framework for DP image synthesis. For example, Ghalebikesabi et al.~\cite{dp-diffusion} apply this pre-training and fine-tuning method using diffusion models. Lyu et al.~\cite{dpldm} propose DP-LDM, which leverages latent diffusion models~\cite{labelembedding} as the synthesizer. Built on DP-LDM, DP-LoRA uses Low-Rank Adapters (LoRA)~\cite{lora}, a parameter-efficient fine-tuning approach, to train latent diffusion models, thus reducing the size of trainable parameters. Li et al.~\cite{li2023PrivImage} introduce PrivImage, which selects a subset of public datasets to align their distribution with sensitive data, resulting in improved synthetic performance.

\end{document}